# Modeling Compressive Instability in Two-Dimensional $Ti_2CO_x$ MXenes

Hossein Darban[1]

*Institute of Fundamental Technological Research, Polish Academy of Sciences, Pawińskiego 5B, 02-106 Warsaw, Poland*

## Abstract

In practical applications, MXenes are often subjected to a variety of loads, including compression. While their mechanical response under different loading conditions, such as tensile loading, has been extensively studied, their compressive instability remains largely unexplored. The compressive and post-buckling behavior of $Ti_2C$ and $Ti_2CO_2$ MXene nanosheets is studied using molecular dynamics (MD) simulations and a nonlocal formulation. The employed interatomic potential is first validated against experimental and density functional theory (DFT) data for structural and mechanical properties. The results indicate that classical continuum mechanics underestimates the buckling strains, whereas the nonlocal formulation adequately captures the observed response. A systematic examination of various defect types up to a defect fraction of 3% reveals that while isolated point defects primarily reduce the critical buckling stress, vacancy clusters significantly alter the buckling mode shapes. Lateral confinement pressure and oxygen surface termination substantially increase the buckling stress. Atomistic analysis reveals opposite stress states in the top and bottom Ti layers due to curvature-induced strain gradients. Under biaxial compression, the nanosheet buckles in a dome-like shape, whereas shear loads produce elliptical deflection modes. The presented findings may stimulate future studies on MXene morphological transformations, such as the development of nanotube, nanoscroll, and folded architectures.

## 1. Introduction

MXenes are a promising family of two-dimensional (2D) transition metal carbides, nitrides, or carbonitrides [1] that have garnered widespread interest due to their remarkable optoelectronic and electrochemical properties, excellent electrical conductivity, hydrophilicity, and robust mechanical properties [2–8]. MXenes have been incorporated into several technologies, including energy storage

[1]- ***Corresponding author:*** *Hossein Darban*; ***E-mail address:*** ***hdarban@ippt.pan.pl***; ***Tel.:*** *(+48) 22 826 12 81*

and conversion systems, catalytic processes, water purification and desalination, electromagnetic interference shielding, communication devices, optical and electronic components, plasmonic systems, sensing and actuation platforms, structural composites, and biomedical applications [7,9–17]. In addition, MXenes can serve as high-performance reinforcements in composite materials for a broad range of applications. For instance, MXene–polymer composites have been explored for biomedical and anticorrosion applications [11,18,19].

The mechanical behavior of MXene nanosheets has been investigated through both experimental and computational approaches; however, experimental data remain limited due to the inherent difficulties of mechanically testing atomically thin 2D materials. Early nanoindentation measurements on single-layer $Ti_3C_2T_x$ MXene using atomic force microscopy reported an elastic modulus of 330 ± 30 GPa [20], notably higher than values obtained for graphene oxide and reduced graphene oxide under similar testing conditions [21,22]. More recently, advances in nanoscale sample fabrication and in-situ tensile testing have enabled direct uniaxial measurements of single-layer $Ti_3C_2T_x$ within an electron microscope, revealing an effective Young's modulus of approximately 480 GPa, elastic strain limits of several percent, and tensile strengths exceeding 15 GPa [23].

As a widely utilized atomistic modeling technique, first-principles simulations have further provided insight into the mechanical properties of MXenes. Density functional theory (DFT) studies on Ti-based MXenes [24], including $Ti_2C$ and $Ti_3C_2$ and their oxygen-functionalized forms, show that these 2D materials exhibit exceptional tensile flexibility. The DFT calculations predicted the detailed stress–strain behavior and revealed that the $Ti_2CO_2$ single-layer can withstand elastic strains exceeding 20%, surpassing even graphene under comparable loading conditions. A recent review [25] further summarizes state-of-the-art DFT and machine-learning methodologies used to investigate MXene structure, properties, and behavior.

While DFT provides high accuracy for small-scale systems involving few atoms, molecular dynamics (MD) simulations are necessary for capturing the mechanical response of larger MXenes. Currently, the investigation of the mechanical properties of MXenes using MD simulations is an active area of research, with reactive force fields (ReaxFF) among the most commonly used interatomic potentials [26]. Because ReaxFF can dynamically model bond formation and breaking, it is particularly well suited to capturing the complex deformation mechanisms that characterize these 2D materials [27–34]. The current state of the art in the mechanical modeling of $Ti_2CT_x$ and $Ti_3C_2T_x$ MXenes includes two widely used ReaxFF parametrizations, originally introduced in [35,36]. A recent study assessed the performance of several interatomic potentials in modeling the structural and elastic properties of titanium carbide and nitride MXenes [37]. The Charge-Optimized Many-Body (COMB3), ReaxFF, and Modified Embedded Atom

Method (MEAM) potentials were evaluated by comparing their predictions of lattice parameters, layer thickness, and elastic constants against available DFT data. The results indicate that the ReaxFF parameterizations, in particular, offer some of the most reliable descriptions of titanium-based MXenes. The recent progress in ReaxFF force field developments for modeling the 2D materials is reviewed in [33].

The ReaxFF force field developed in [36] has been particularly effective for the mechanical modeling of MXenes. For example, it was used in [32] to demonstrate that oxygen functionalization strongly affects the tensile response of $Ti_2C$ MXene, such as its stiffness, strength, and failure strain. Similarly, the work in [34] employed this potential to show that surface terminations significantly influence the elastic and fracture behavior of Ti-based MXenes. The force field has also been applied to study the friction and failure of $Ti_3C_2T_x$ under sliding [27], the reduction of fracture toughness due to atomic-scale defects in $Ti_2C$ [28], and nanoindentation responses of $Ti_2CO_2$ and $Ti_3C_2O_2$, where point defects were found to lower the Young's modulus [29]. Additionally, the study in [30] used the same ReaxFF parameterization to demonstrate that surface termination engineering can effectively tailor the tensile properties of $Ti_3C_2T_x$ MXenes. Other classical MD potentials have also been employed to investigate the tensile and fracture behavior of various MXenes [6,31,38,39].

In practical applications such as flexible electronics, wearable devices, and protective coatings, MXenes are often subjected to compressive, bending, or shear stresses during repeated mechanical deformation [40–43]. Moreover, during the fabrication of MXene–polymer composites, nanosheets can experience compressive loading due to polymerization-induced matrix shrinkage, a phenomenon also observed in carbon-nanotube-based composites [44]. Consequently, MXene nanosheets may buckle, wrinkle, detach, or fail. Therefore, studying compressive and shear responses is essential for designing mechanically robust MXene-based materials and predicting their performance under operating conditions. However, such fundamental knowledge remains largely absent from the current literature. Most experimental and numerical studies on MXene nanosheets have focused on their tensile or flexural properties, while investigations of their compressive and shear behavior remain scarce. One notable exception is the work reported in [45], where in-situ uniaxial compression experiments on accordion-like multilayer $Ti_2CT_x$ and $Ti_3C_2T_x$ revealed pronounced anisotropy and highly nonlinear deformation under both in-plane and out-of-plane loading. Using SEM-assisted nanoindentation, the study revealed progressive structural failure, characterized by bending, buckling, tearing, shearing, kinking, and delamination, as the compressive load increased.

The scarcity of studies on the compressive behavior of MXenes is especially apparent in comparison with other 2D materials such as graphene, where multiple MD investigations have rigorously characterized both compression- and shear-induced buckling mechanisms [46–49]. The present study aims to address this scientific gap by conducting a series of reactive molecular dynamics simulations using the well-established ReaxFF force field developed in [36]. The simulations provide novel insights into instability mechanisms and systematically examine the effects of defects, confinement pressure, and surface terminations on the buckling response. Furthermore, the results demonstrate that a nonclassical continuum mechanics formulation is required to accurately capture the observed buckling behavior of MXene nanosheets.

The remainder of this article is organized as follows: Section 2 details the MD simulation framework and provides the parameters necessary to reproduce the reported results. Section 3 presents original findings on the buckling behavior of $Ti_2C$ and $Ti_2CO_2$ MXenes, beginning with a validation of the ReaxFF force field. Its accuracy in predicting in-plane lattice parameter, monolayer thickness, and bond lengths, as well as Young's modulus, is benchmarked against available DFT and experimental data from the literature. Furthermore, we demonstrate the limitations of classical formulations in capturing size-dependent buckling and successfully implement a stress-driven nonlocal model as a robust alternative. Finally, Section 4 summarizes the findings and offers concluding remarks.

## 2. Molecular Dynamics Simulations

Molecular dynamics (MD) simulations are employed in this work to investigate the mechanical response and stability of $Ti_2C$ MXene under compression and shear. Both uniaxial and biaxial compression cases are considered. The effects of strain rate, lateral confinement pressure, isolated point and vacancy cluster defects, and $O_2$ surface termination on the response of $Ti_2C$ MXene are studied.

The MD simulations are conducted using the Large Scale Atomic/Molecular Massively Parallel Simulator (LAMMPS, 29 Aug 2024) [50–52] and the ReaxFF force field developed in [36]. The obtained results are visualized using the Open Visualization Tool (OVITO) [53]. The simulations involve $Ti_2C$ and $Ti_2CO_2$ MXene nanosheets with an in-plane dimension of $14 \times 14$ nm$^2$, as shown in Fig. 1. The initial atomic configurations are constructed by replicating the unit cells obtained from [54] along in-plane directions in OVITO. Simulations are performed at 1 K to suppress local atomic thermal vibrations and isolate the intrinsic mechanical response of the 2D MXene nanosheets. A timestep of 0.2 fs is used for all simulations in this study, unless otherwise stated.

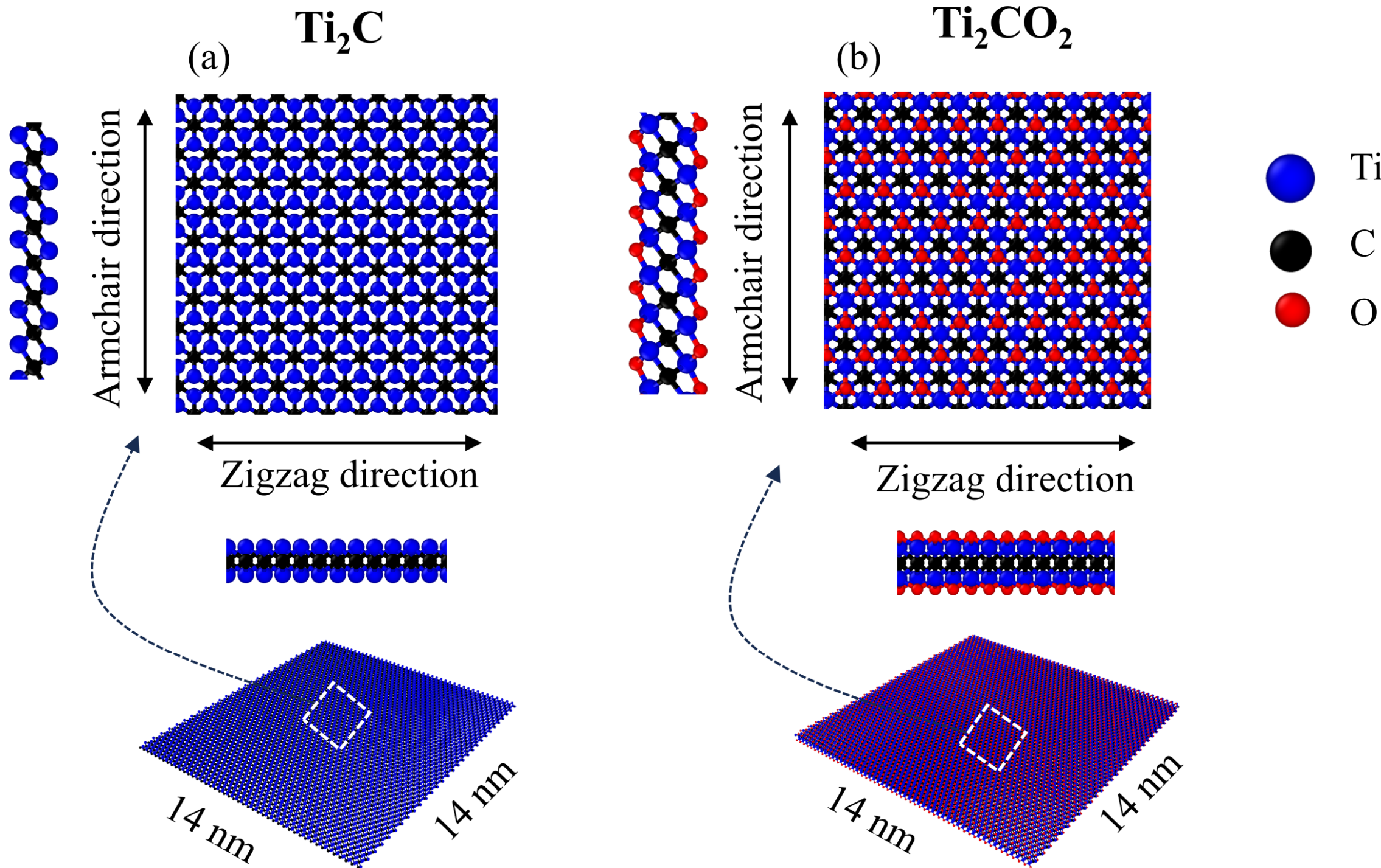

**Fig. 1** Initial atomic configuration of (a) $Ti_2C$ and (b) $Ti_2CO_2$ MXene nanosheets.

Three-dimensional simulations are run with non-periodic boundary conditions to model the behavior of a free-standing MXene nanosheet. After assigning the initial coordinates of atoms, the system undergoes an energy minimization process in which atomic positions are iteratively adjusted until both the energy and force fall below predefined tolerances of $1\times10^{-10}$ (dimensionless) and $1\times10^{-10}$ (kcal/mol)/Å, respectively. This procedure relaxes the atomic configuration to an energy minimum, thereby eliminating residual stresses. The minimization was conducted in LAMMPS using the conjugate gradient (CG) algorithm with the Polak–Ribiere (PR) formulation. This particular CG variant is recognized as highly efficient for diverse minimization tasks in atomistic modeling. The velocity form of the Stoermer-Verlet time integration algorithm (velocity-Verlet) is used [55].

Following minimization, atomic velocities at the studied temperature of 1 K are defined. Then, the system is equilibrated for 300 ps at the temperature of 1 K using the NVT ensemble, where the temperature is controlled using a Nose-Hoover thermostat. During equilibration, a lateral confinement pressure is applied to the MXene nanosheet to preserve its flat geometry and facilitate the onset of the plate-like first buckling mode (see Fig. 2 (d)). This is implemented by introducing repulsive walls positioned 1.6 Å above and below the MXene nanosheet, each interacting with the atoms within a cutoff distance of 1.5 Å. The interaction energy between a wall and an atom at a distance $r$ follows the 9–3

Lennard-Jones (LJ) potential, $E=\varepsilon\left(2/15(\sigma/r)^{9}-(\sigma/r)^{3}\right)$ with $\varepsilon=0.1$ kcal/mol and $\sigma=1.74749$ Å. These parameter values are selected to maintain the flat geometry of the MXene nanosheet without exerting excessive repulsion. They also ensure that the force on the atoms is zero at the cutoff distance and increases progressively as atoms approach the wall. The energy, temperature, and internal forces of the system are computed and monitored during equilibration to ensure the system is properly equilibrated before mechanical testing (see Fig. 2(a)-(c)). The in-plane force components are negligible during relaxation. In contrast, the out-of-plane force $F_z$ exhibits oscillations between ±1 *nN* (see Fig. 2(c)), arising from interactions of the nanosheet with the repulsive wall potentials. The confinement pressure is released at the end of the relaxation stage, before deformation is applied.

The deformation at the edge of the nanosheet was applied by displacing a few atomic layers in the boundary region, with a thickness of 1 nm, at a constant velocity corresponding to the target strain rate and under the NVT ensemble. Note that the portion of the MXene nanosheet free to move during deformation had an effective size of 14 nm × 14 nm. At each timestep, the strain is computed as the ratio of the applied displacement to the initial length of the nanosheet along the loading direction (14 nm).

The stress is obtained directly from LAMMPS output using the virial formulation. To convert the virial stress from stress × volume units (LAMMPS output) to true stress, an effective volume must be defined for the 2D MXene sheet, which requires assigning a thickness to the nanosheet. As in other 2D materials, this thickness can be defined either as an effective shell thickness derived from bending rigidity or as a geometric thickness corresponding to the distance between the uppermost and lowermost atomic layers. First-principles calculations reported in [56] demonstrate that these two definitions yield very similar values for $Ti_2C$ and $Ti_2CO_2$ MXenes. In the present study, the geometric thickness definition is adopted.

It is noted that a small residual stress is present at the onset of external loading. This residual stress arises from the simulation procedure. As described above, the nanosheets are initially relaxed in the presence of repulsive wall potentials; following relaxation, the walls are removed prior to the application of deformation. This transition, combined with the finite size of the system and the boundary constraints imposed during loading, may introduce a slight pre-stress in the structure. In most cases, the magnitude of this residual stress is small. Importantly, its presence does not affect the qualitative trends of the mechanical response reported in this study.

To quantify the potential influence of the empirically chosen 9-3 LJ wall parameters ($\varepsilon$, $\sigma$) on the pre-stress state, a sensitivity analysis was performed. Two additional simulations were conducted wherein the position of the LJ walls and their interaction parameters were systematically varied. In the first test case (wall placed further from the MXene), the parameters were set to $\varepsilon = 0.2$, $\sigma = 2.91248$ Å, with a cutoff of 2.5 Å and the wall positioned at 2.6 Å. In the second test case (wall placed closer), the parameters were set to $\varepsilon = 0.05$, $\sigma = 0.58250$ Å, with a cutoff of 0.5 Å and the wall positioned at 0.6 Å. For both modified configurations, the parameters and cutoff distances were calibrated to ensure that the interaction force smoothly decayed to exactly zero at the cutoff boundary. An analysis of the pre-stress state before compression revealed that varying the wall distance and interaction parameters in this manner had a negligible effect on the internal pre-stress of the system.

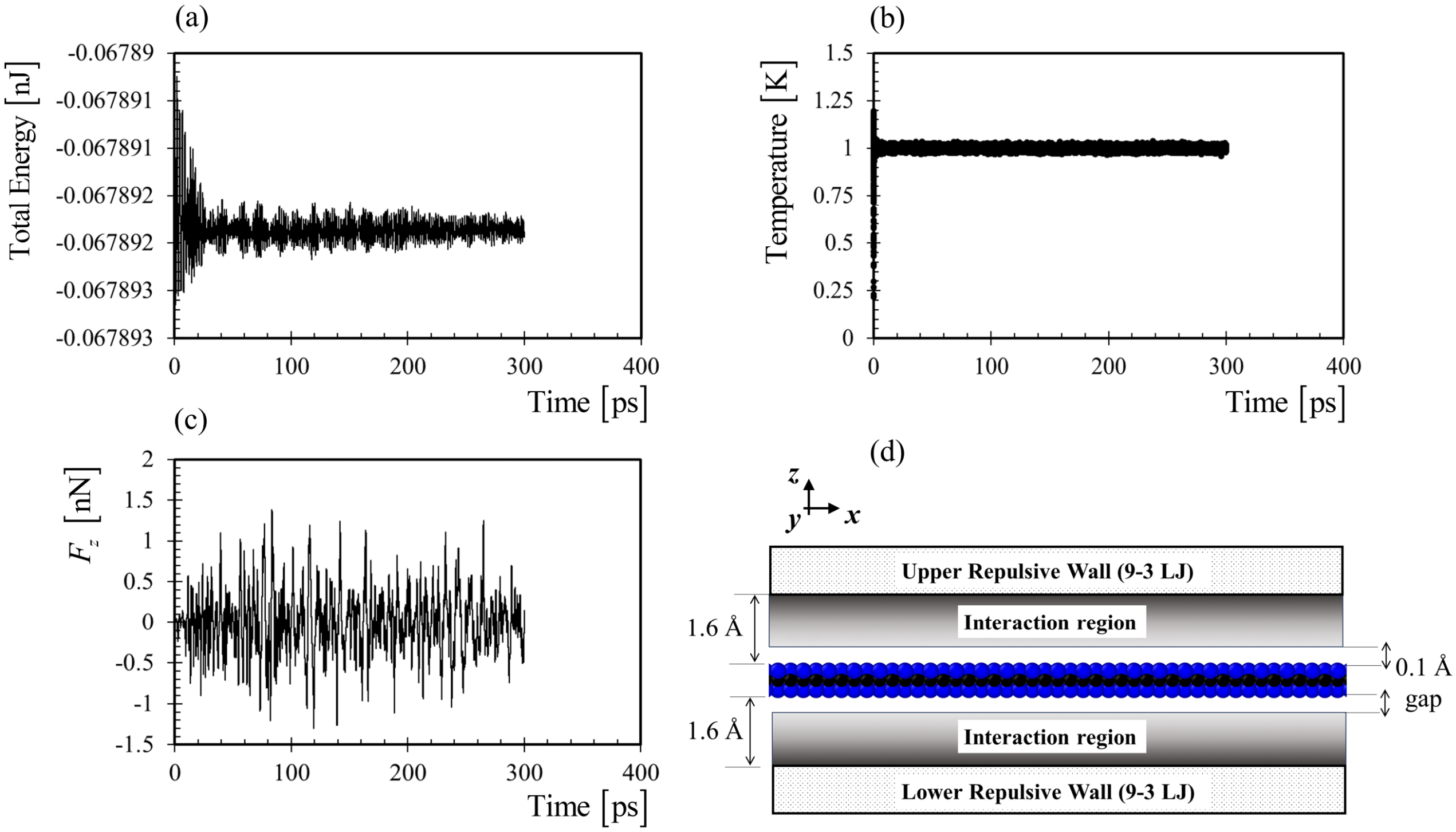

**Fig. 2** Thermomechanical variables during equilibrium: (a) total energy, (b) temperature, and (c) out-of-plane force vs. time. (d) Schematic illustrating the lateral confinement pressure applied in the out-of-plane direction during equilibrium.

Although the problem could, in principle, be investigated using molecular statics (MS) simulations at 0 K, our analysis revealed a pronounced sensitivity of the response to the applied displacement rate. Accurately capturing the buckling response of MXene nanosheets within an MS framework would require exceedingly small displacement increments accompanied by extensive relaxation at each loading step, resulting in a prohibitively high computational cost. In contrast, the MD approach adopted in this work offers a more practical solution and inherently enables subsequent studies of temperature effects.

## 3. Results and Discussions

The computations in this study were carried out on the state-of-the-art supercomputer Ares, hosted by the Academic Computer Centre CYFRONET AGH in Kraków, Poland. One computing node, equipped with 48 cores provided by dual Intel Xeon Platinum 8268 processors (2.90 GHz) and 192 GB of RAM, was employed. The total wall-clock time for the simulations varied depending on the number of atoms in the model and the total simulation duration. For example, the simulations presented in Section 3.2, which investigate the strain rate effect, required approximately 40 hours to complete, whereas those described in Section 3.6 on the post-buckling behavior took about one week.

### 3.1. Model Validation

The accuracy of the MD simulations strongly depends on the reliability of the employed interatomic potential. The ReaxFF force field adopted in this study [36] has been previously validated in the literature for different titanium-based MXenes, including contributions by the author (see Ref. [57]). For completeness, the predicted in-plane lattice parameter, monolayer thickness, bond lengths, and Young's modulus of $Ti_2C$ and $Ti_2CO_2$ MXenes are depicted in Fig. 3, showing good agreement with available data from density functional theory (DFT) calculations and experiments reported in the literature.

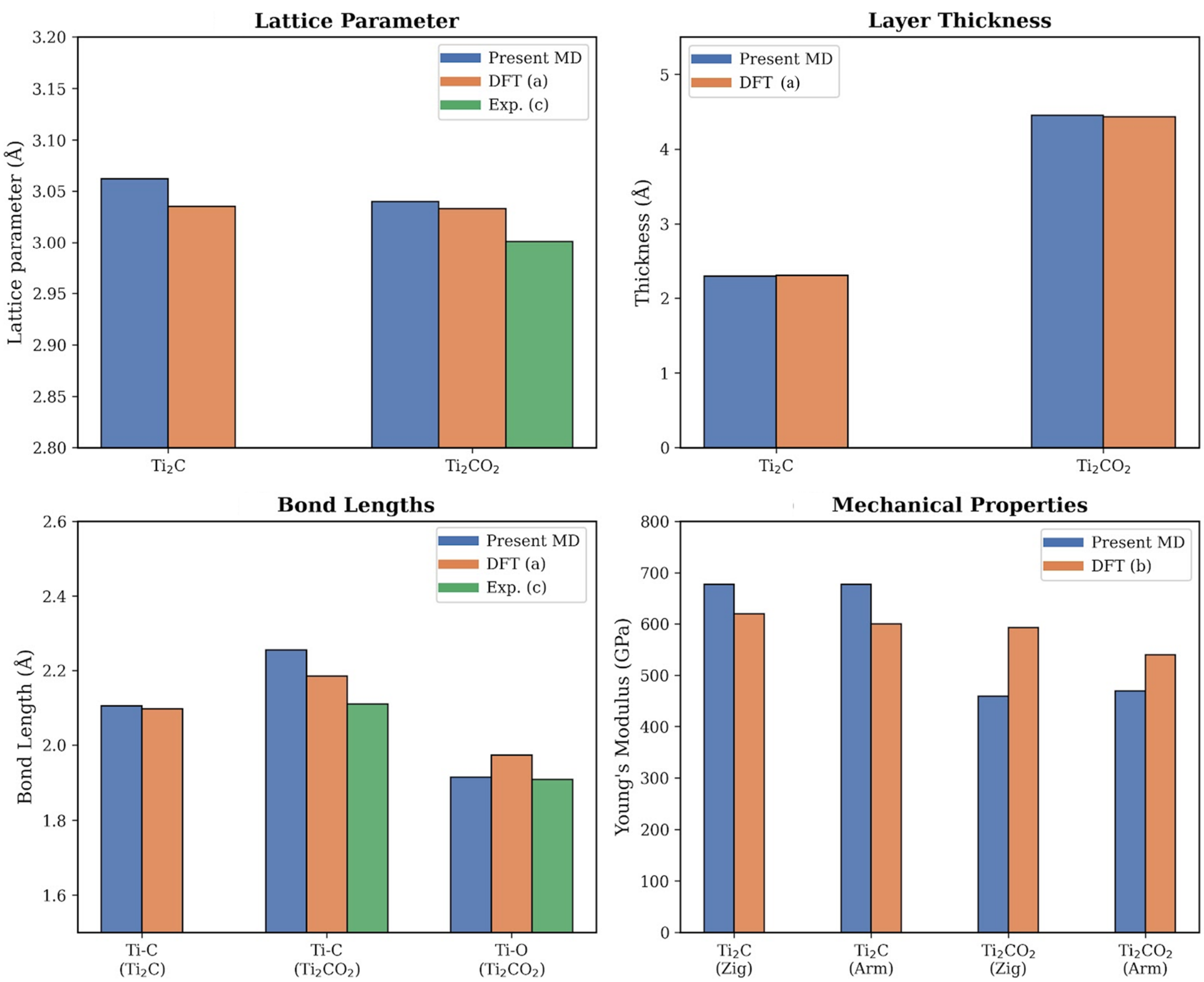

**Fig. 3** Comparison of calculated structural and mechanical parameters of $Ti_2C$ and $Ti_2CO_2$ MXenes with literature values. The present MD results show good consistency with DFT calculations and experimental data for in-plane lattice parameter, layer thickness, atomic bond lengths, and Young's modulus. The literature data are taken from (a) [58], (b) [24], and (c) [59].

For structural properties, equilibrium configurations were determined using a 20×20 supercell subjected to conjugate gradient energy minimization. The system was allowed to relax under zero in-plane pressure to identify optimal lattice parameters and layer thicknesses. Atomic bond lengths were subsequently extracted from the first peak of the radial distribution function (RDF). The in-plane Young's modulus was evaluated through quasi-static uniaxial tensile tests on 50×10 nm sheets along the armchair and zigzag directions. Following initial relaxation, incremental displacements (2 fm) were applied to the sheet boundaries. The Young's modulus was derived from the linear slope of the virial stress–strain response within a 0.3% strain range, using the previously determined monolayer thicknesses for volume normalization.

### 3.2. Strain Rate Effect

The strain rates accessible in MD simulations are indeed several orders of magnitude higher than those typically associated with real-world applications. This limitation, which may influence the predicted buckling loads and strains, is intrinsic to atomistic time integration and cannot be avoided without compromising computational feasibility. However, the objective of the present study is to elucidate the fundamental deformation and buckling mechanisms at the atomic scale rather than to reproduce application-specific loading rates quantitatively. The simulations offer mechanistic insight that is expected to be qualitatively transferable to lower strain-rate conditions.

The stress–strain curves are shown in Fig. 4 for a 14×14 $nm^2$ $Ti_2C$ MXene nanosheet subjected to five different strain rates. The curves corresponding to compression along the armchair and zigzag directions. From the initial linear portion of the stress–strain curves, the Young's moduli along the armchair and zigzag directions are determined to be 655 GPa and 590 GPa, respectively. These values differ slightly from those presented in Fig. 3, as they correspond to compressive loading and a different simulation setup. However, these values are still in good agreement with previously reported data from first-principles calculations for the tensile response of $Ti_2C$ MXenes, which indicate Young's moduli of approximately 600 GPa and 620 GPa along the armchair and zigzag directions, respectively [24].

MXene nanosheets subjected to higher compression rates exhibit buckling at higher stress and strain levels and display more unstable post-buckling behavior, characterized by stress oscillations of greater amplitude. As the strain rate decreases, the buckling stress tends to converge, and for the three lowest strain rates, the buckling stress values become nearly identical. The curves in Fig. 4(c) compare the compressive response of the MXene nanosheet along the armchair and zigzag directions at the strain rate of $0.725\times10^6$ $s^{-1}$. As shown, the nanosheet sustains higher stresses when compressed along the armchair direction. For compression along the armchair direction, the $Ti_2C$ MXene nanosheet exhibits buckling at a stress and strain of approximately 1 GPa and 0.0015, whereas buckling occurs at about 0.6 GPa and 0.0006 when compressed along the zigzag direction. The predicted buckling strain of 0.0006 for the zigzag-oriented loading case is affected by residual stresses present at zero strain.

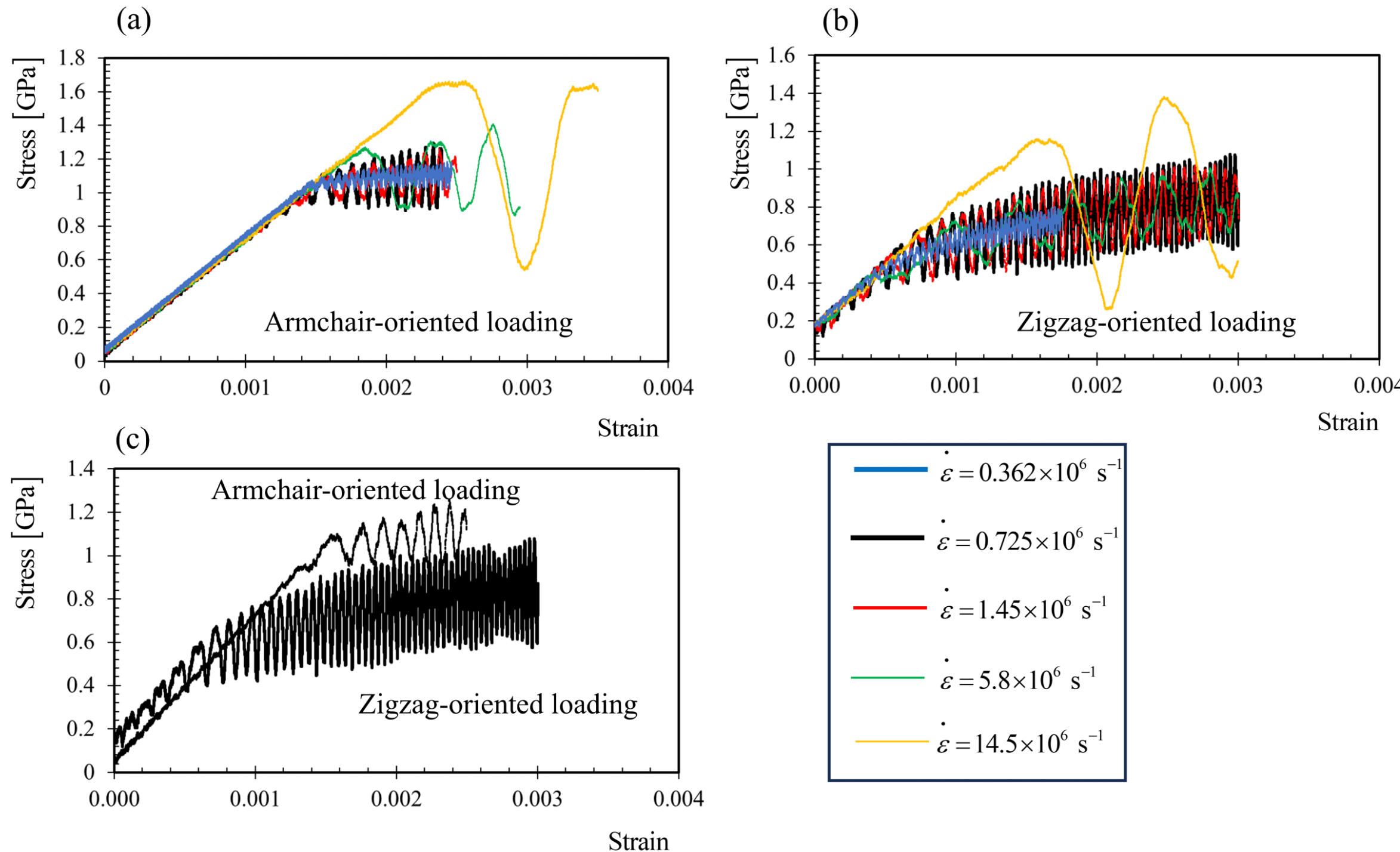

**Fig. 4** Stress-strain curves of $Ti_2C$ MXene nanosheets with an in-plane dimension of 14×14 nm$^2$ under compression along (a) the armchair direction and (b) the zigzag direction at different strain rates. (c) Comparison of the stress-strain curves under armchair- and zigzag-oriented loading at a strain rate of $0.725\times10^{6}$ s$^{-1}$ and timestep of 0.2 fs.

The adopted simulation setup induces the plate-like first buckling mode in $Ti_2C$ MXene nanosheets. The corresponding buckling mode shapes at different applied strains are shown in Fig. 5 for loadings along both the armchair and zigzag directions. At a strain of 0.00117, the nanosheet already exhibits buckling under zigzag compression, while it remains flat under loading along the armchair direction. With increasing strain, the maximum out-of-plane deflection, occurring at the midsection of the nanosheet, grows progressively, with the deflection under armchair-oriented loading consistently smaller than that under zigzag-oriented loading. In both loading directions, the buckling mode shapes display the inherent symmetry associated with mode I deformation.

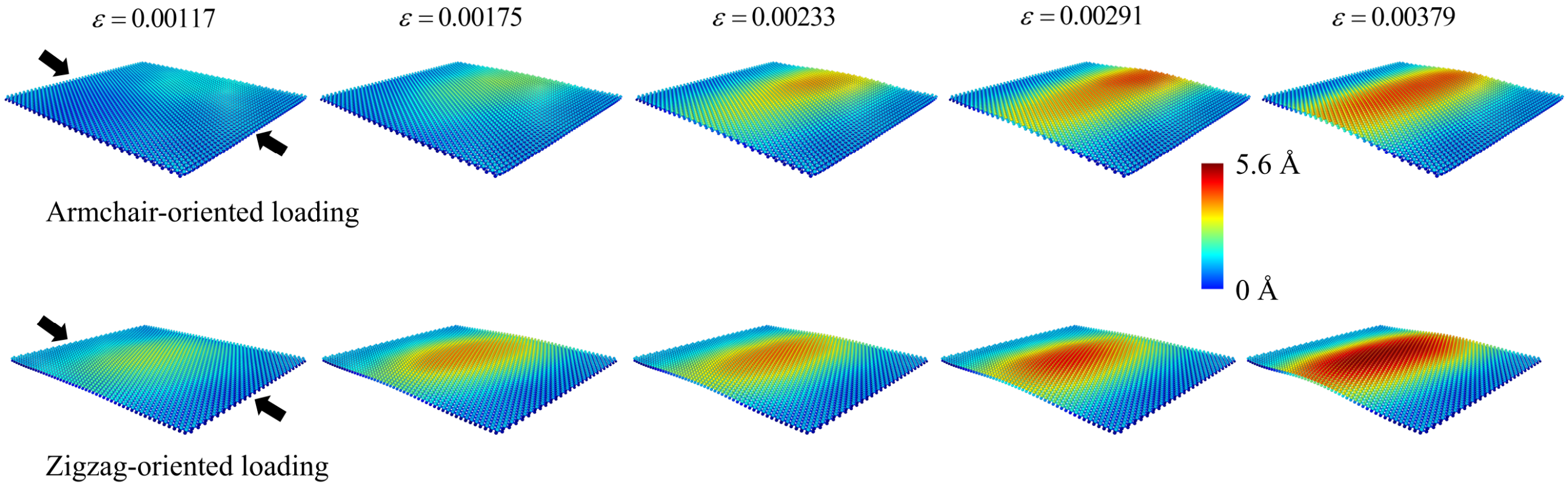

**Fig. 5** Buckling mode shapes of $Ti_2C$ MXene nanosheets with an in-plane dimension of 14×14 nm$^2$ under progressive compression along the armchair and the zigzag directions at the strain rate of $1.45\times10^6$ s$^{-1}$ and timestep of 0.2 fs. Color maps represent the out-of-plane deflections.

## 3.3. Effect of Defect

MXene nanosheets may contain vacancy defects, either inherited from their MAX phase precursors or generated during chemical etching. The presence of such defects reduces the stiffness of MXenes, making them more susceptible to mechanical instabilities under compression. In this section, the effect of vacancy defects on the compressive response of $Ti_2C$ MXene nanosheets is investigated.

Section 3.3.1 examines MXene nanosheets with randomly distributed single-point defects, without distinguishing between Ti and C atoms. This scenario mimics the isolated defects typically observed in MXenes synthesized using low- to medium-concentration hydrofluoric (HF) acid [60]. In Section 3.3.2, we systematically separate the Ti and C point defects to isolate and understand the individual contributions of each atom type. In Section 3.3.3, we investigate the effect of Ti vacancy clusters on the buckling behavior of the nanosheets. This clustered defect scenario represents the more severe structural degradation characteristic of MXenes etched with high HF concentrations [60].

### *3.3.1 Random Point Defects*

Fig. 6 presents the stress–strain curves and associated buckling configurations for nanosheets with in-plane dimensions of 14×14 nm$^2$, compressed along the armchair and zigzag directions with varying point defect fractions. The simulations were conducted at a strain rate of $1.45\times10^6$ s$^{-1}$, using a timestep of 0.2 fs. The desired defect fraction is introduced by randomly removing the corresponding number of atoms from the simulation box, without distinguishing between Ti and C atoms.

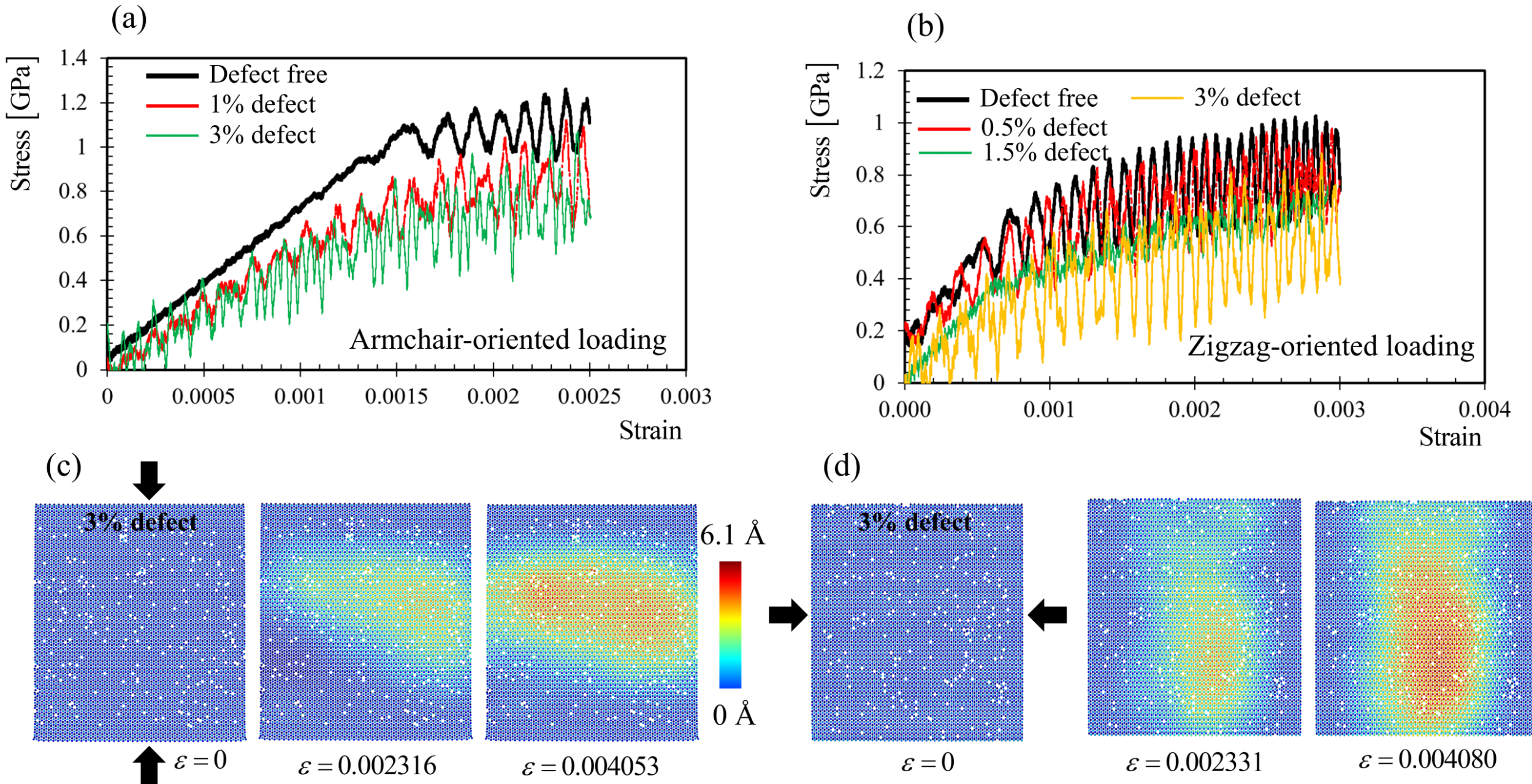

**Fig. 6** Stress-strain curves of $Ti_2C$ MXene nanosheets with an in-plane dimension of 14×14 nm$^2$ under compression along (a) the armchair direction and (b) the zigzag direction for different percentages of random point defects at the strain rate of $1.45\times10^6$ s$^{-1}$ and timestep of 0.2 fs. Buckling mode shapes related to compression along (c) the armchair and (d) the zigzag directions. Color maps represent the out-of-plane deflections.

The stress–strain curves in Fig. 6(a) and (b) reveal a clear dependence of the compressive response of $Ti_2C$ MXene nanosheets on the vacancy defects. As the defect percentage increases, a noticeable reduction in peak stress is observed for both loading directions. In addition to lowering the critical buckling stress, vacancy defects also alter the post-buckling behavior. Higher defect fractions lead to increased amplitude of stress oscillations in the post-buckling regime, indicating reduced stability and greater susceptibility to local structural collapse. This observation is further supported by the buckling configurations shown in Fig. 6(c) and (d), where larger out-of-plane deflections are evident in regions with a higher density of vacancy defects. The vacancy defect exhibits no significant influence on the buckling strain.

The reduction in buckling stress with increasing defect fraction can be attributed to the degradation of the nanosheet stiffness. The presence of defects disrupts the atomic bonding network, leading to a decrease in the effective elastic modulus (see Fig. 6(a)). As a result, the structure becomes more compliant and less capable of sustaining high stresses before instability. This trend is consistent with previous MD studies, which have shown that defects reduce the Young's modulus of MXenes [29].

To assess whether the influence of vacancy defects on the compressive response of MXene nanosheets remains consistent across different strain rates, additional simulations were performed. These simulations follow the same setup as in Fig. 6(a), considering both defect free and 3% defected nanosheets, but at a lower strain rate of $0.725\times10^{6}$ s$^{-1}$. The resulting stress–strain curves are shown in Fig. 7. As observed, the overall trends remain consistent with those obtained at a higher strain rate of $1.45\times10^{6}$ s$^{-1}$, indicating that the effect of vacancy defects is insensitive to the strain rate within the range considered.

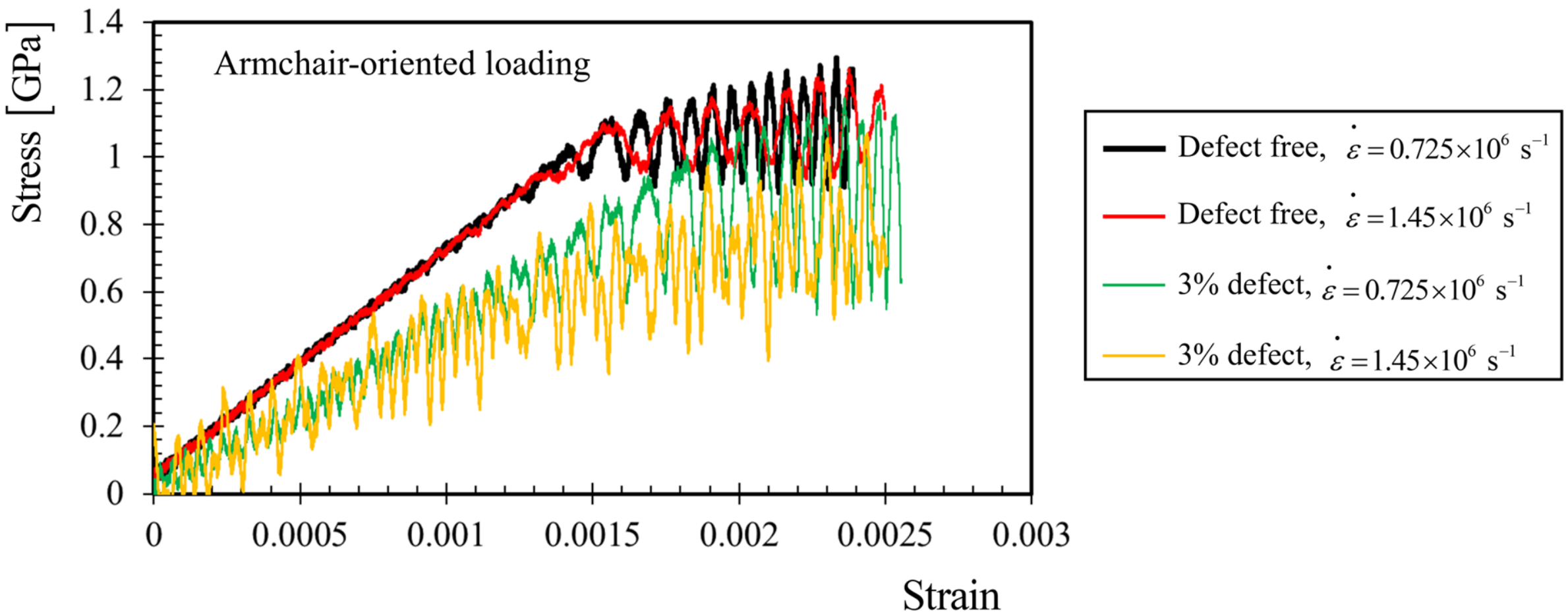

**Fig. 7** Stress-strain curves of $Ti_2C$ MXene nanosheets with an in-plane dimension of $14\times14$ nm$^2$ under compression along the armchair direction for defect free and 3% defected nanosheets at two strain rates of $1.45\times10^{6}$ s$^{-1}$ and $0.7255\times10^{6}$ s$^{-1}$. The timestep in simulations is 0.2 fs.

*3.3.2 Ti or C Point Defects*

This section examines the buckling response of $Ti_2C$ MXenes containing isolated point defects located in either the Ti or C atomic layers. Defects are introduced into the pristine MXene via a random spatial distribution, subject to a minimum separation distance of 3.5 Å to prevent vacancy cluster formation. For Ti point defects, no symmetry constraint is enforced to balance the number of defects between the two Ti layers. Consequently, the resulting MXene configurations with Ti defects exhibit an asymmetric distribution of vacancies across the upper and lower layers.

The stress–strain curves are presented in Fig. 8(a) and (b). Isolated Ti point defects considerably influence the compressive response of $Ti_2C$ MXenes. As the Ti defect fraction increases, the nanosheet progressively weakens and buckles at lower critical stresses. In contrast, the isolated C point defects considered here do not have a pronounced effect on the buckling response. This indicates that, within the investigated range of C defect fractions (up to 3%), the stiffness of the MXene nanosheet remains largely insensitive to the presence of C vacancies. This observation is consistent with the results reported in [29] based on MD simulations of monolayer MXene indentation.

The out-of-plane deflections shown in Fig. 8(c) and (d) are consistent with the trends observed in the stress–strain response. In the presence of isolated Ti defects, for a given strain level, the deflection increases with defect fraction due to the enhanced compliance of the nanosheets. This trend is not observed for isolated C defects, as the stiffness remains effectively unchanged within the studied defect range. Consequently, at a fixed strain, the out-of-plane deflection does not necessarily increase with increasing C defect fraction. For example, the nanosheet containing 3% C defects exhibits a lower deflection at a strain of 0.00348 compared to the cases with 1% and 2% defects.

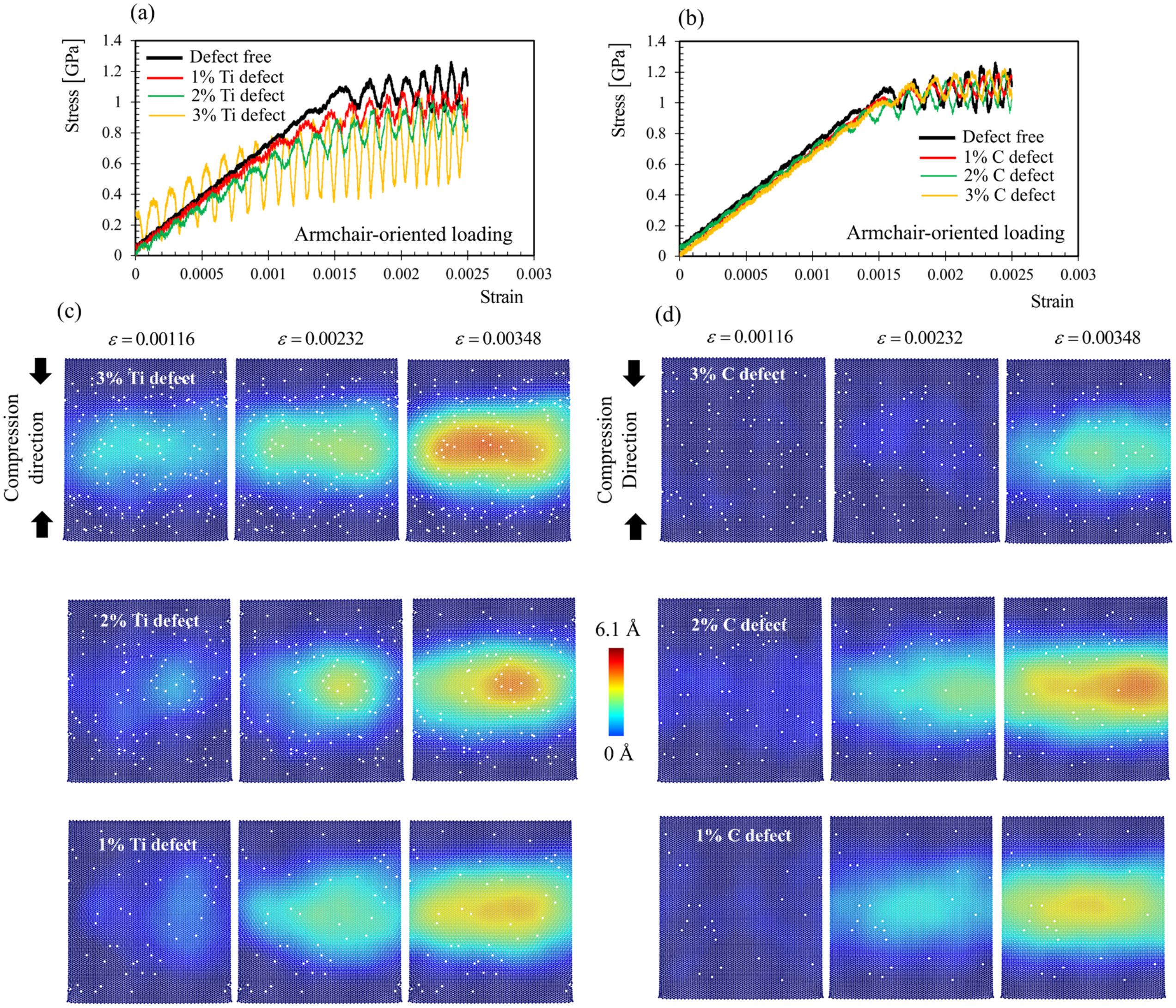

**Fig. 8** Stress-strain curves of $Ti_2C$ MXene nanosheets with an in-plane dimension of 14×14 nm$^2$ under compression along the armchair direction for different percentages of isolated (a) Ti and (b) C point defects at the strain rate of $1.45\times10^6$ $s^{-1}$ and timestep of 0.2 fs. Buckling mode shapes related to MXenes with (c) Ti and (d) C point defects. Color maps represent the out-of-plane deflections.

*3.3.3 Ti Vacancy Clusters*

The effect of Ti vacancy clusters on the buckling behavior of MXene nanosheets is examined in this section. The stress–strain curves and out-of-plane deflections shown in Fig. 9 correspond to $Ti_2C$ MXene nanosheets containing 3% Ti defects arranged in clusters of 2, 4, and 5 vacancies. Defect clusters are introduced into pristine MXenes with a random spatial distribution, subject to a minimum separation distance of 3.5 Å to prevent the formation of vacancy clusters with a higher number of vacancies.

A primary observation is that the nanosheets no longer exhibit classical thin-plate buckling behavior. Instead, the deformation is characterized by localized out-of-plane deflections in regions with a higher density of vacancy clusters. These clustered defects introduce localized compliance, which promotes the development of such non-uniform deformations. As a result, the nanosheets lose their initially flat configuration at relatively early stages of compression. Furthermore, the intensity of these local deflections increases with the number of vacancies per cluster.

All three stress–strain curves display a common trend: the initial stiffness is reduced compared to that of defect free MXene nanosheets. Compared to the case of 3% isolated Ti point defects (see Fig. 8(a)), nanosheets with clusters of two and four vacancies exhibit similar maximum stresses in the stress–strain response, whereas the configuration with five-vacancy clusters exhibits a noticeably higher maximum stress. This behavior may also be influenced by the spatial distribution of the clusters, which is not explicitly accounted for in the present analysis.

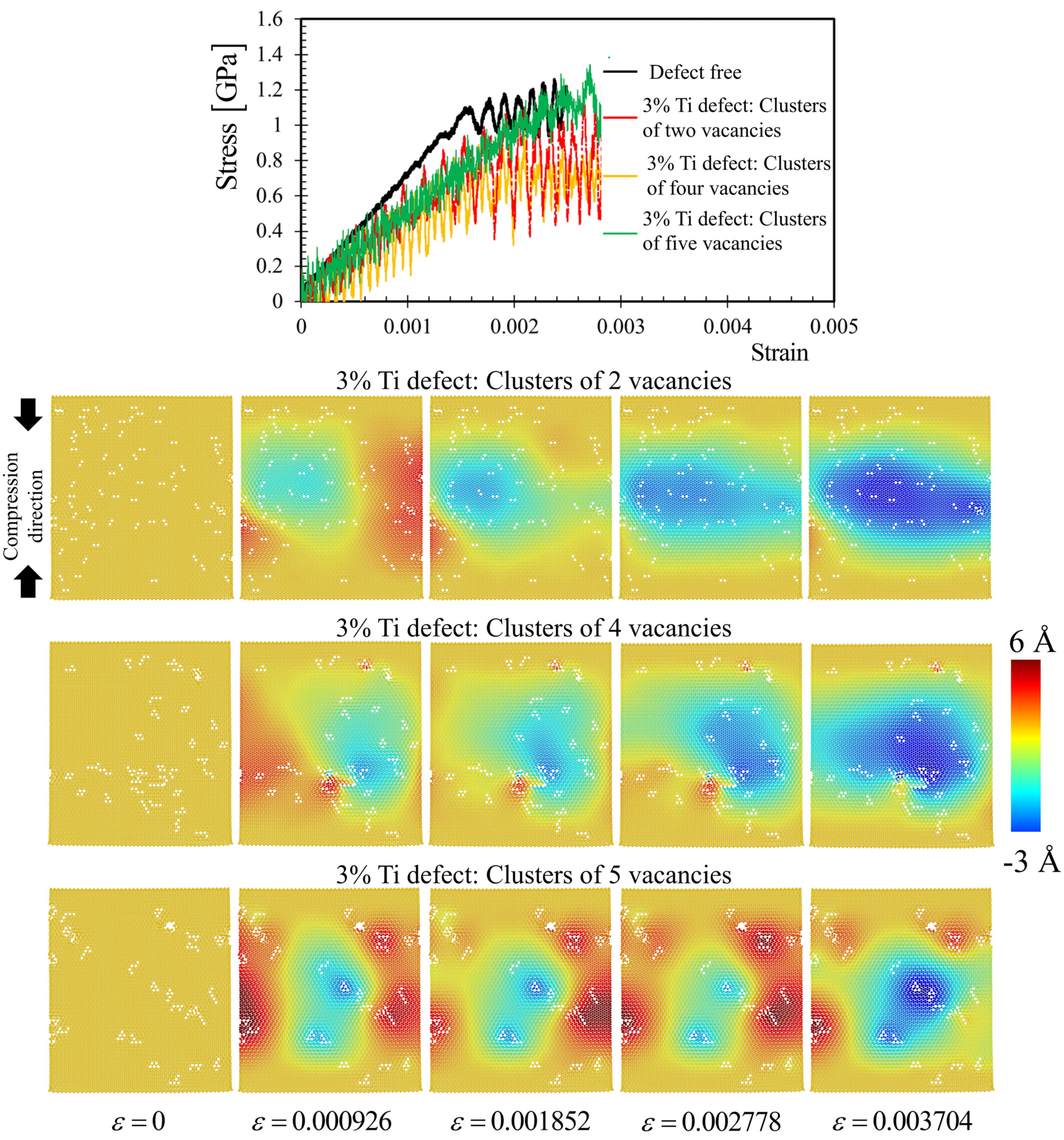

**Fig. 9** Stress-strain curves of $Ti_2C$ MXene nanosheets with an in-plane dimension of 14×14 $nm^2$ under compression along the armchair direction for different percentages of Ti vacancy clusters. The strain rate is $1.45\times10^6$ $s^{-1}$ and timestep 0.2 fs. Buckling mode shapes are also shown at different strains. Color maps represent the out-of-plane deflections.

### 3.4. Lateral Confinement Pressure Effect

In this section, we examine how confinement pressure influences the compressive response of $Ti_2C$ MXene nanosheets. The lateral confinement pressure during the deformation stage is introduced by positioning two layers of harmonic springs above and below the MXene nanosheet with an initial gap of 0.1 Å. This interaction is analogous to a Winkler elastic foundation and is implemented in LAMMPS using the "wall/region" command. Simulations are then performed for various spring stiffness values, and the corresponding stress–strain responses are shown in Fig. 10. These results refer to $Ti_2C$ MXene nanosheets with an in-plane dimension of $14\times14$ nm$^2$ subjected to compression along the armchair direction at a strain rate of $5.8\times10^6$ s$^{-1}$ using a timestep of 0.2 fs. The specific stiffness values, $k = 0.0002$ to 0.001 (kcal/mol)/Angstrom$^2$, were selected to ensure a smooth potential energy surface for a parametric analysis of the confinement effect on the buckling response. In a realistic, material-specific scenario, the true effective stiffness is not an independent parameter; rather, an approximation can be obtained from the mechanical properties of the surrounding matrix as well as the specific geometry of the interacting phases [61].

As shown by the stress–strain curves, applying a lateral confinement pressure increases the buckling stress and strain, reduces the amplitude of post-buckling stress oscillations, and has no effect on the slope of the curves. For example, a $Ti_2C$ MXene embedded in a medium with a stiffness of 0.001 (kcal/mol)/Angstrom$^2$ exhibits nearly twice the buckling stress of a free-standing MXene.

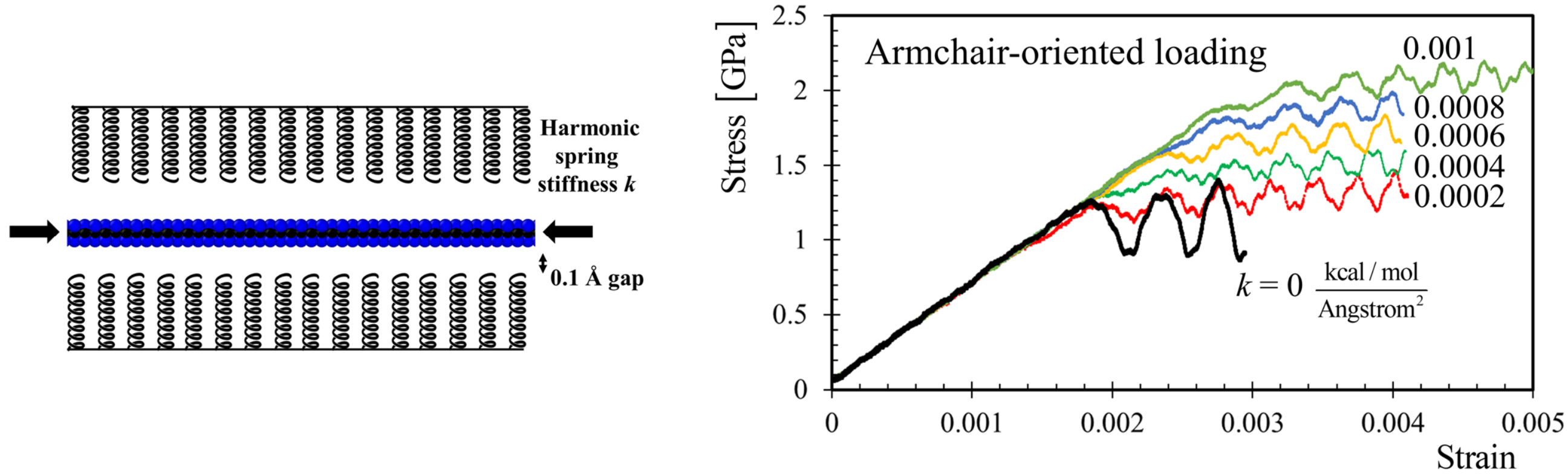

**Fig. 10** Stress-strain curves of $Ti_2C$ MXene nanosheets with an in-plane dimension of $14\times14$ nm$^2$ under compression along the armchair direction for different lateral confinement pressures at the strain rate of $5.8\times10^6$ s$^{-1}$ and timestep of 0.2 fs.

The increase in buckling stress and strain with increasing spring stiffness can be attributed to the additional constraint provided by the surrounding interaction, which effectively increases the overall stiffness of the system. This behavior is consistent with well-established findings in the literature, where the presence of an elastic foundation is known to increase the critical buckling load of plates, including at the nanoscale. For instance, an increase in the buckling stress of nanoplates supported by an elastic medium has been reported in [62] using analytical formulations based on nonlocal and surface elasticity theories.

The spring stiffness $k$ effectively governs the degree of mechanical constraint imposed on the nanosheet. In the context of an MXene-reinforced composite, $k$ can be interpreted as a representation of the stiffness of the surrounding matrix. Specifically, lower values of $k$ correspond to a more compliant matrix, whereas higher values are associated with a stiffer embedding medium. Based on the results presented in Fig. 10, it can be inferred that an MXene flake embedded in a stiffer matrix can sustain higher compressive loads and strains prior to buckling, thereby delaying the onset of instability and potential interfacial detachment between the reinforcement and the matrix.

To further examine the role of lateral confinement pressure on the buckling response, a complementary set of simulations was carried out at a reduced strain rate. The analysis follows the same framework as in Fig. 10, comparing unsupported nanosheets ($k$ = 0) with laterally supported ones ($k$ = 0.001 (kcal/mol)/Angstrom$^2$ ) at a strain rate of $1.45\times10^6$ s$^{-1}$ . The corresponding stress–strain responses are reported in Fig. 11. The comparison reveals no qualitative deviation from the behavior observed at the higher strain rate of $5.8\times10^6$ s$^{-1}$ , suggesting that the impact of lateral confinement pressure on buckling remains effectively unchanged over the investigated strain-rate range.

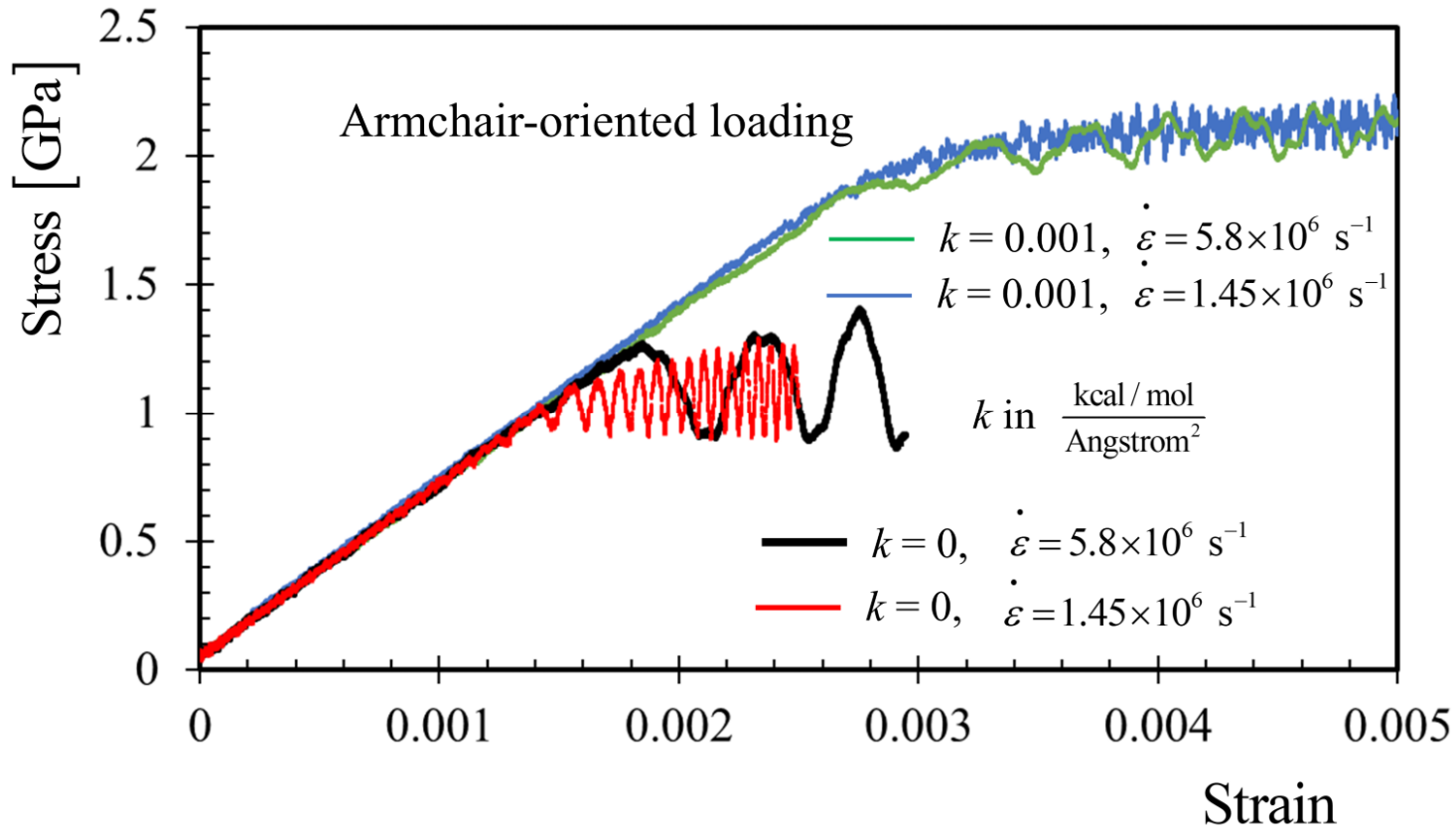

**Fig. 11** Stress-strain curves of $Ti_2C$ MXene nanosheets with an in-plane dimension of 14×14 nm$^2$ under compression along the armchair direction for unsupported and supported nanosheets at the strain rates of $5.8\times10^6$ and $1.45\times10^6$ s$^{-1}$ and timestep of 0.2 fs.

### 3.5. Effect of $O_2$ Surface Termination

MXenes typically exhibit surface terminations such as –O, –OH, or –F as a result of their synthesis routes, providing opportunities to tailor their properties. In this section, we examine how –O termination influences the compressive behavior of $Ti_2C$ MXene nanosheets. For this purpose, $Ti_2C$ and $Ti_2CO_2$ nanosheets with in-plane dimensions of 14×14 nm$^2$ are simulated under compression along both the armchair and zigzag directions at a strain rate of $5.8\times10^6$ s$^{-1}$ using a timestep of 0.2 fs. The obtained stress-strain curves are illustrated in Fig. 12.

The –O surface termination leads to a substantial increase in the buckling stress of the MXene nanosheets. For the geometry considered here, the buckling stress rises from roughly 1 GPa to nearly 3.5 GPa. This enhancement is primarily attributed to the increased effective thickness introduced by the termination groups, which in turn raises the second moment of area and thus the bending stiffness of the nanosheet. Another notable outcome is that –O termination reduces the anisotropy between the responses under armchair and zigzag loading. Furthermore, the Young's moduli of $Ti_2C$ nanosheets, 655 GPa along the armchair direction and 590 GPa along the zigzag direction, decrease to approximately 480 GPa in $Ti_2CO_2$. This reduction aligns with first-principles predictions reported in [24]. However, the Young's modulus values for $Ti_2CO_2$ provided in [24] are higher, reported as 540 GPa (armchair) and 593 GPa (zigzag).

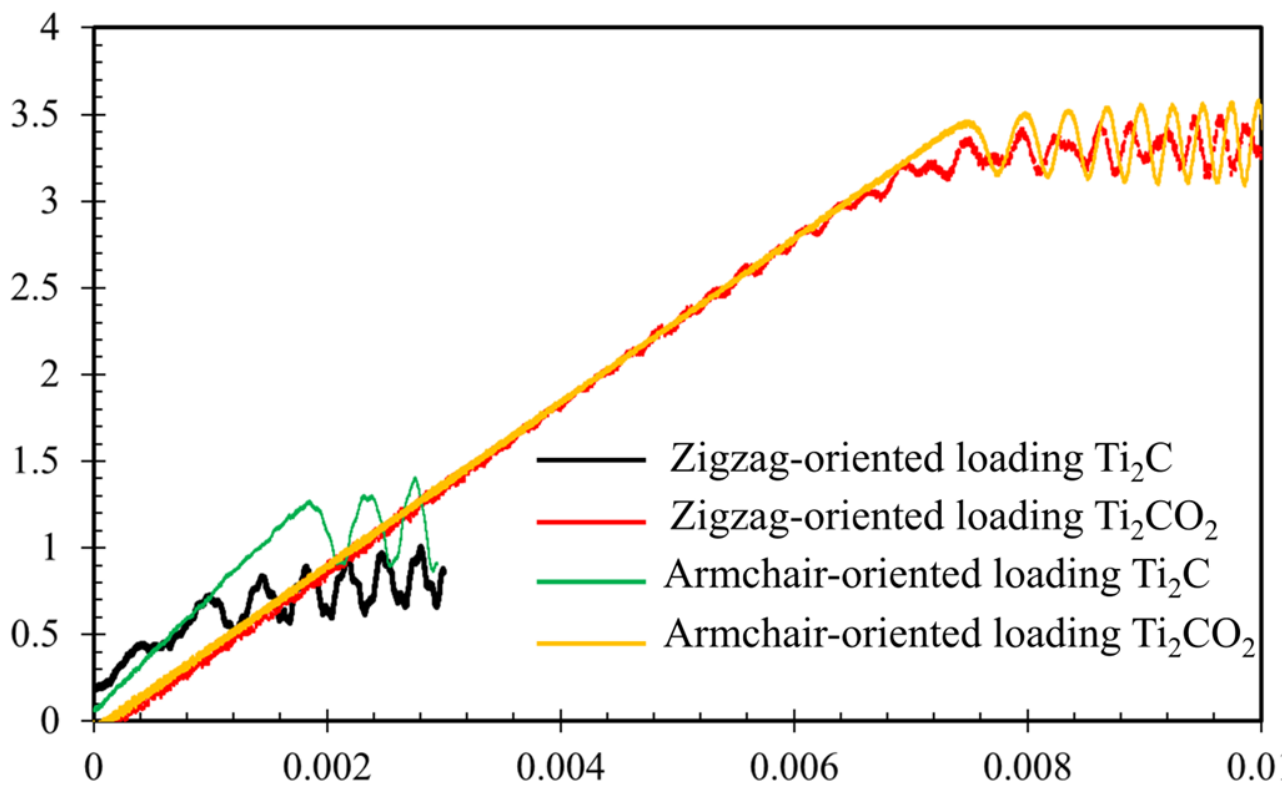

**Fig. 12** Stress-strain curves of $Ti_2C$ and $Ti_2CO_2$ MXene nanosheets with an in-plane dimension of 14×14 nm$^2$ under compression along the armchair and the zigzag directions at the strain rate of $5.8\times10^6$ s$^{-1}$ and timestep of 0.2 fs.

### 3.6. Post-Buckling

To examine the post-buckling behavior of $Ti_2C$ nanosheets, additional simulations are performed at a higher strain rate of $29\times10^6$ s$^{-1}$ to maintain a feasible computational cost. Both defect free and 2% defective $Ti_2C$ MXene nanosheets (randomly distributed point defects similar to the cases in Section 3.3.1) with an in-plane dimension of 14×14 nm$^2$ are subjected to progressive compression along the armchair direction. The resulting buckling mode shapes are presented in Fig. 13. The simulations are extended to strains exceeding 0.26, during which no fracture is observed. The presence of 2 % random point defects does not noticeably alter the global buckling patterns, even at these large strains, indicating that their influence remains localized.

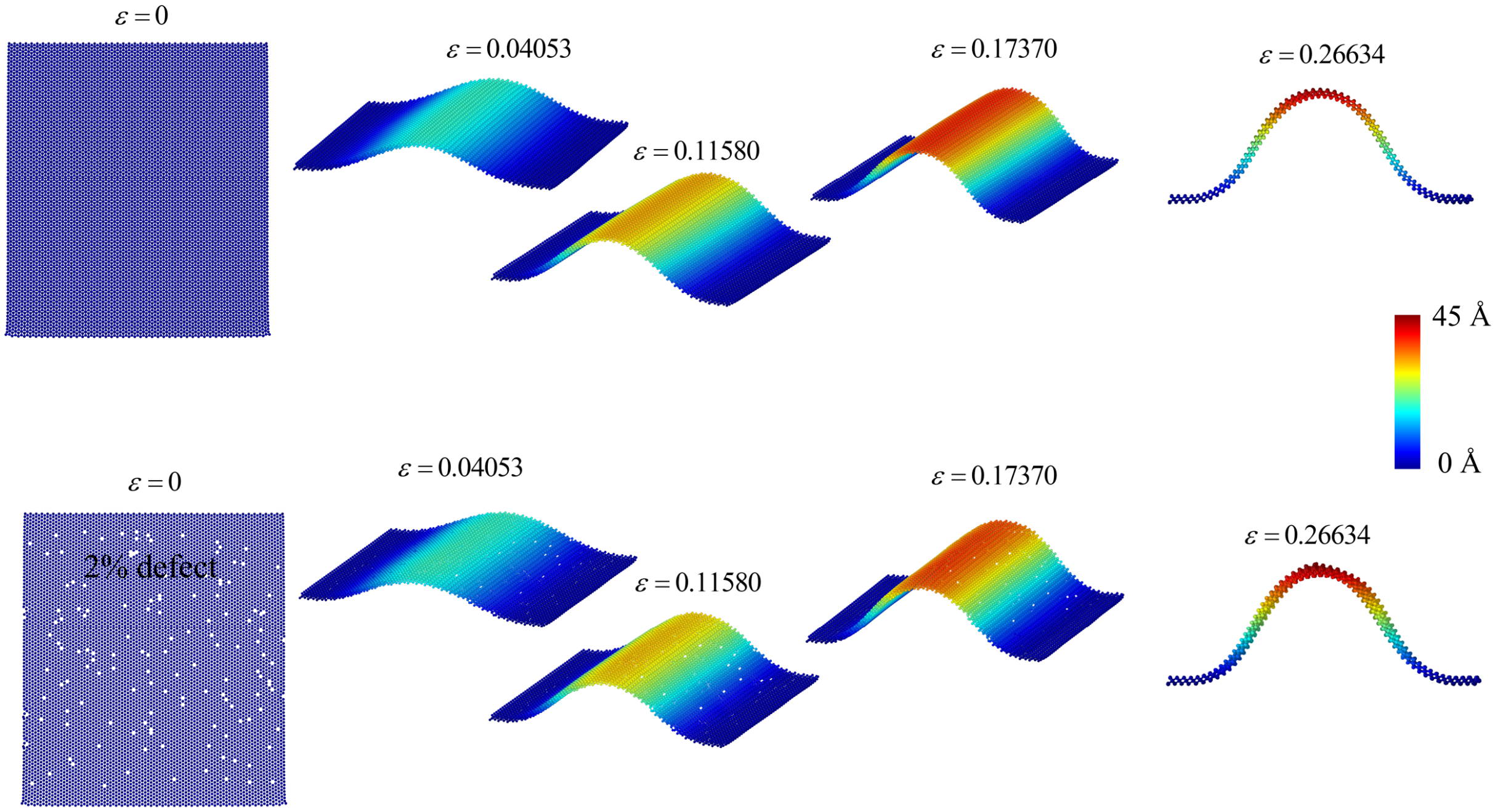

**Fig. 13** Buckling mode shapes of defect free and 2% defective (randomly distributed point defects) $Ti_2C$ MXene nanosheets with an in-plane dimension of 14×14 nm$^2$ under progressive compression along the armchair direction at the strain rate of $29\times10^6$ s$^{-1}$ and timestep of 0.2 fs. Color maps represent the out-of-plane deflections.

Although the $Ti_2C$ MXene nanosheets do not fracture under buckling, even at high strain levels, their lattice structure undergoes notable adjustments to accommodate the strain distribution through the thickness. For example, in the central region of the nanosheet shown in Fig. 13, the curvature bends downward. This implies that the upper Ti layer experiences tensile strain along the loading direction, whereas the lower Ti layer is subjected to compression. This behavior is illustrated in Fig. 14, which displays the evolution of Ti–Ti and C–C bond lengths during loading; bonds longer than 3.08 Å are omitted for clarity.

As shown, the upper Ti layer develops bond lengths exceeding 3.08 Å along the loading direction in the central region, while the bond lengths perpendicular to the load remain mostly unchanged. In contrast, the lower Ti layer shows shortened bonds in the same central region. Moving toward the first and last quarters of the nanosheet, the trend reverses: the lower Ti layer exhibits increased bond lengths along the loading direction, whereas the upper layer's bond lengths decrease. The C–C bond lengths are less sensitive to deformation compared to the Ti–Ti bonds throughout the loading process. However, at higher strain levels, the C–C bond lengths also exceed 3.08 Å along the loading direction in the left, central, and right regions of the nanosheet.

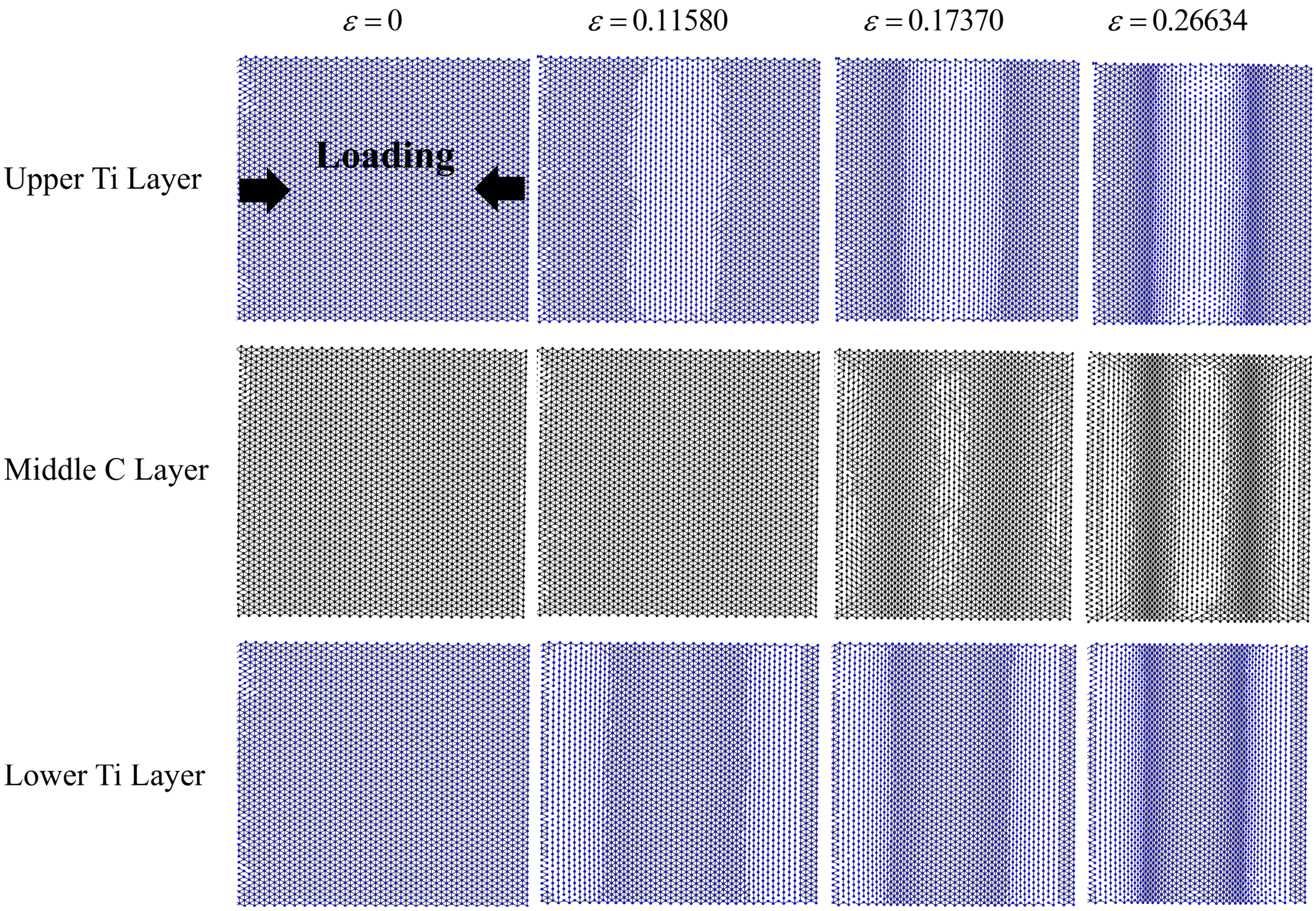

**Fig. 14** Analysis of Ti-Ti and C-C bond lengths in a $Ti_2C$ MXene nanosheet with an in-plane dimension of $14\times14$ nm$^2$ under progressive compression along the armchair direction at the strain rate of $29\times10^6$ s$^{-1}$ and timestep of 0.2 fs. Bonds with lengths higher than 3.08 Å are not shown.

To examine how the lattice structure evolves under loading, the variations in Ti–Ti and C–C bond lengths and bond angles are tracked within a small representative cell located at the center of the nanosheet. This cell consists of three atoms in each of the Ti–C–Ti layers. As illustrated in Fig. 15, the three atoms in each layer form an equilateral triangle at zero strain.

For the upper Ti layer, the bond lengths are labeled $U_1$, $U_2$, and $U_3$, with the corresponding angles denoted as $\theta_{12}$, $\theta_{13}$, and $\theta_{23}$. $U_1$ represents the bond length perpendicular to the loading direction and remains nearly constant throughout deformation. In contrast, $U_2$ and $U_3$ increase almost equally as the applied strain increases, confirming that the upper Ti layer experiences tensile deformation in the central region of the nanosheet. For instance, at the strain of 0.25, the bond length $U_2$ is stretched by 0.16 Å. Consistent with the changes in the bond lengths, $\theta_{12}$ and $\theta_{13}$ increase, whereas $\theta_{23}$ decreases. In the lower Ti layer, $L_1$, the bond length perpendicular to the loading direction also remains mostly unchanged. However, $L_2$ and $L_3$ decrease with increasing strain, indicating compression of the lower-layer Ti atoms. The corresponding bond angles adjust in a manner consistent with this compressive deformation. The C–C bonds and angles in the middle carbon layer exhibit only minor changes during loading, reflecting the relatively lower sensitivity of this layer to the imposed deformation.

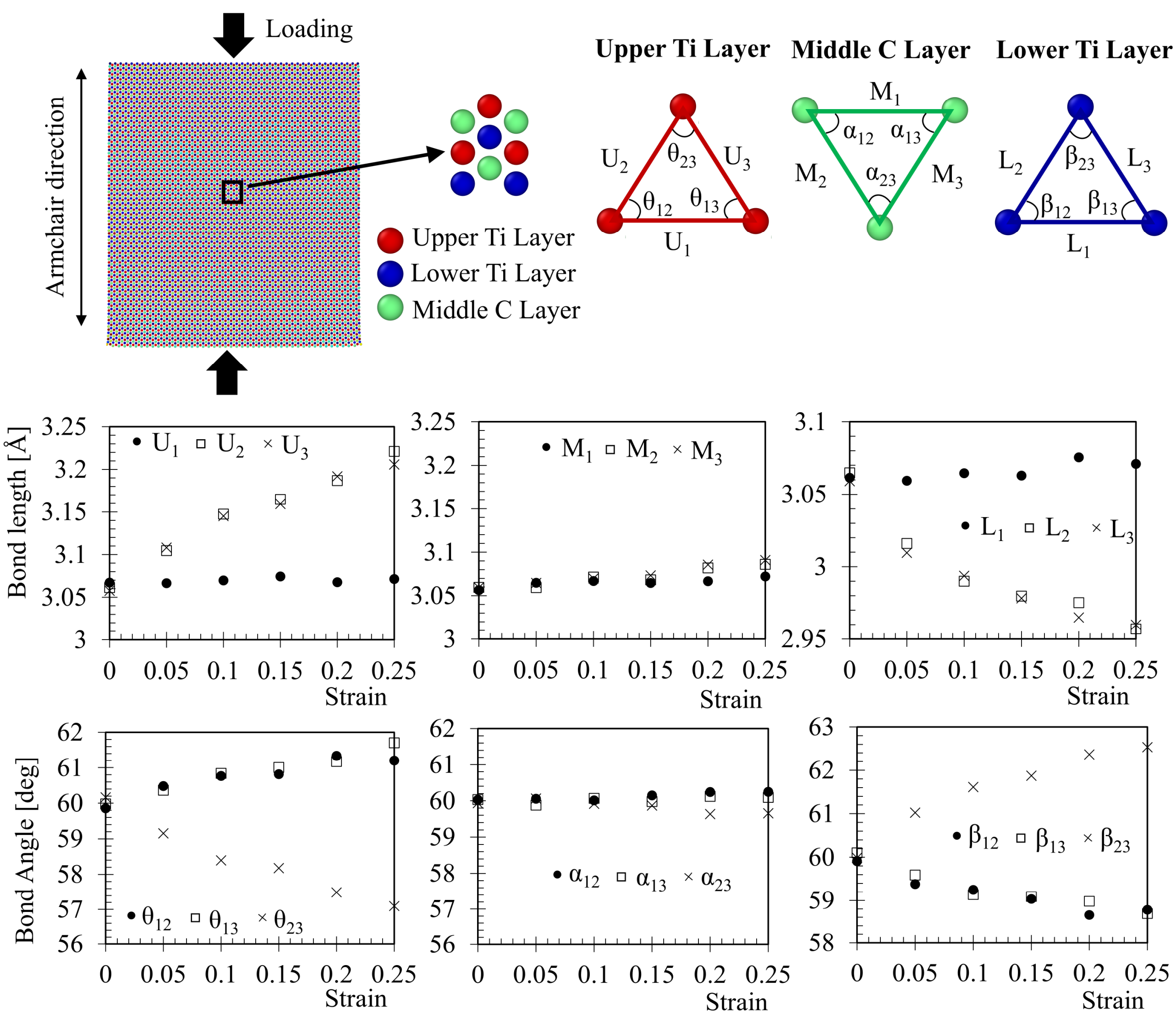

**Fig. 15** Variations of Ti-Ti and C-C bond lengths and angles in the center of $Ti_2C$ MXene nanosheet with an in-plane dimension of 14×14 $nm^2$ under progressive compression along the armchair direction at the strain rate of $29\times10^6$ $s^{-1}$ and timestep of 0.2 fs.

To assess how oxygen surface termination influences the post-buckling behavior, we performed additional simulations on $Ti_2C$ and $Ti_2CO_2$ MXene nanosheets with in-plane dimensions of 14×14 $nm^2$. The nanosheets were compressed along both the armchair and zigzag directions at an elevated strain rate of $2.85\times10^8$ $s^{-1}$ using a timestep of 1 fs. These higher values were chosen to keep the computational cost manageable. The corresponding stress–strain responses are presented in Fig. 16. The $Ti_2C$ nanosheets exhibit a smooth bending deformation without any signs of fracture, even at strains as large as 0.35. In contrast, the $Ti_2CO_2$ nanosheets fractured completely in the middle section. The fracture strain is notably higher when the $Ti_2CO_2$ nanosheet is loaded along the zigzag direction compared to loading along the armchair direction.

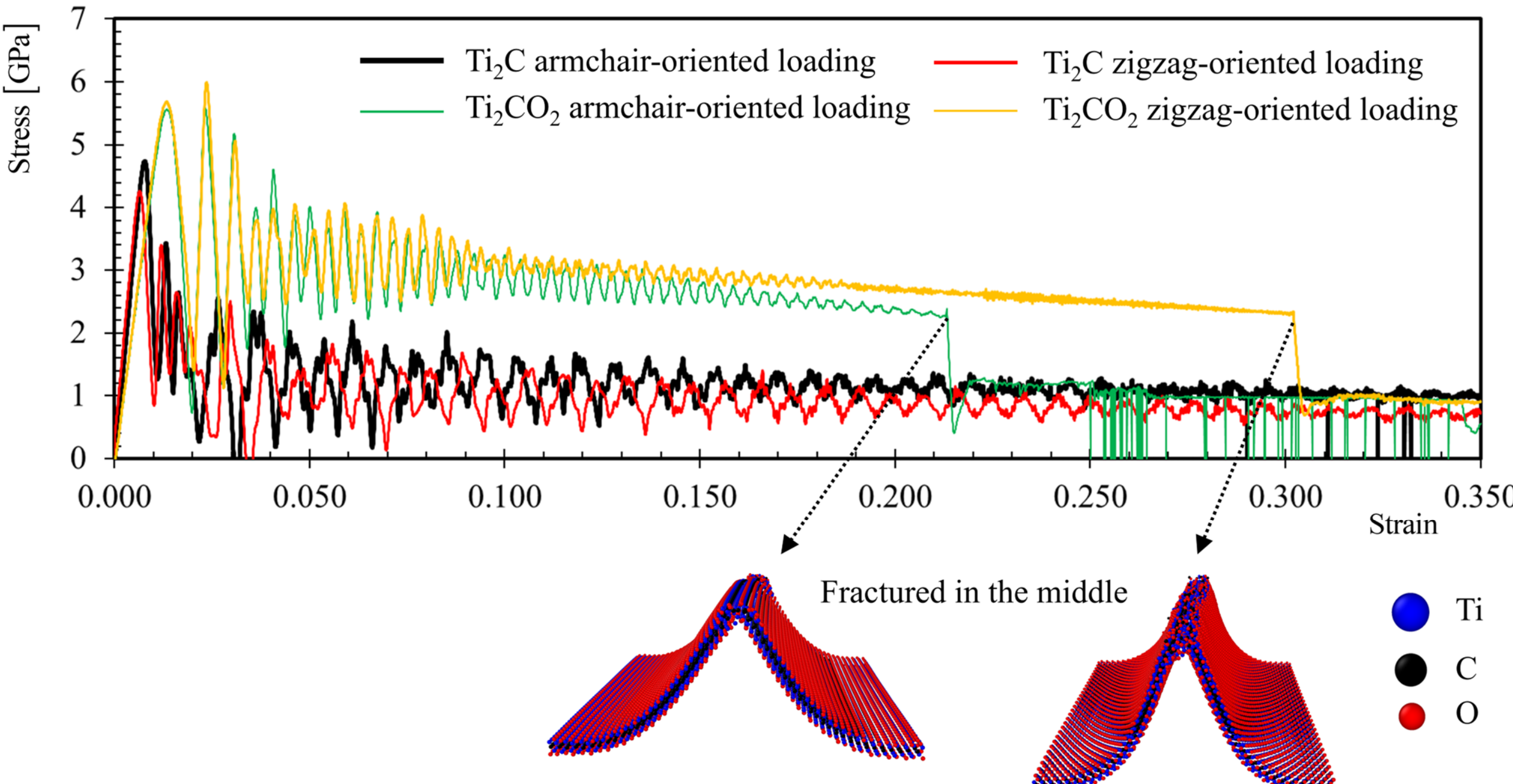

**Fig. 16** Stress-strain curves of $Ti_2C$ and $Ti_2CO_2$ MXene nanosheets with an in-plane dimension of 14×14 $nm^2$ under compression along the armchair and the zigzag directions at the strain rate of $2.85\times10^8$ $s^{-1}$ and timestep of 1 fs.

### 3.7. Biaxial and Shear Buckling

The buckling behavior of a $Ti_2C$ MXene nanosheet with in-plane dimensions of 14×14 $nm^2$ under biaxial compression is examined in this section. The nanosheet is compressed uniformly along both in-plane directions at a strain rate of $1.45\times10^{6}$ $s^{-1}$. The resulting buckling mode shapes, shown in Fig. 17(a), display maximum out-of-plane deflection at the nanosheet's center. The iso-deformation contours (lines connecting atoms with equal deflection) form nearly concentric circles, indicating a symmetric buckling pattern. The corresponding stress–strain curves are provided in Fig. 17(b). The stress–strain curves along the armchair and zigzag directions nearly coincide, consistent with the symmetric deformation mode. The resulting dome-like buckling pattern further reflects the similarity in in-plane stiffness of the two directions and the uniformity of the applied load.

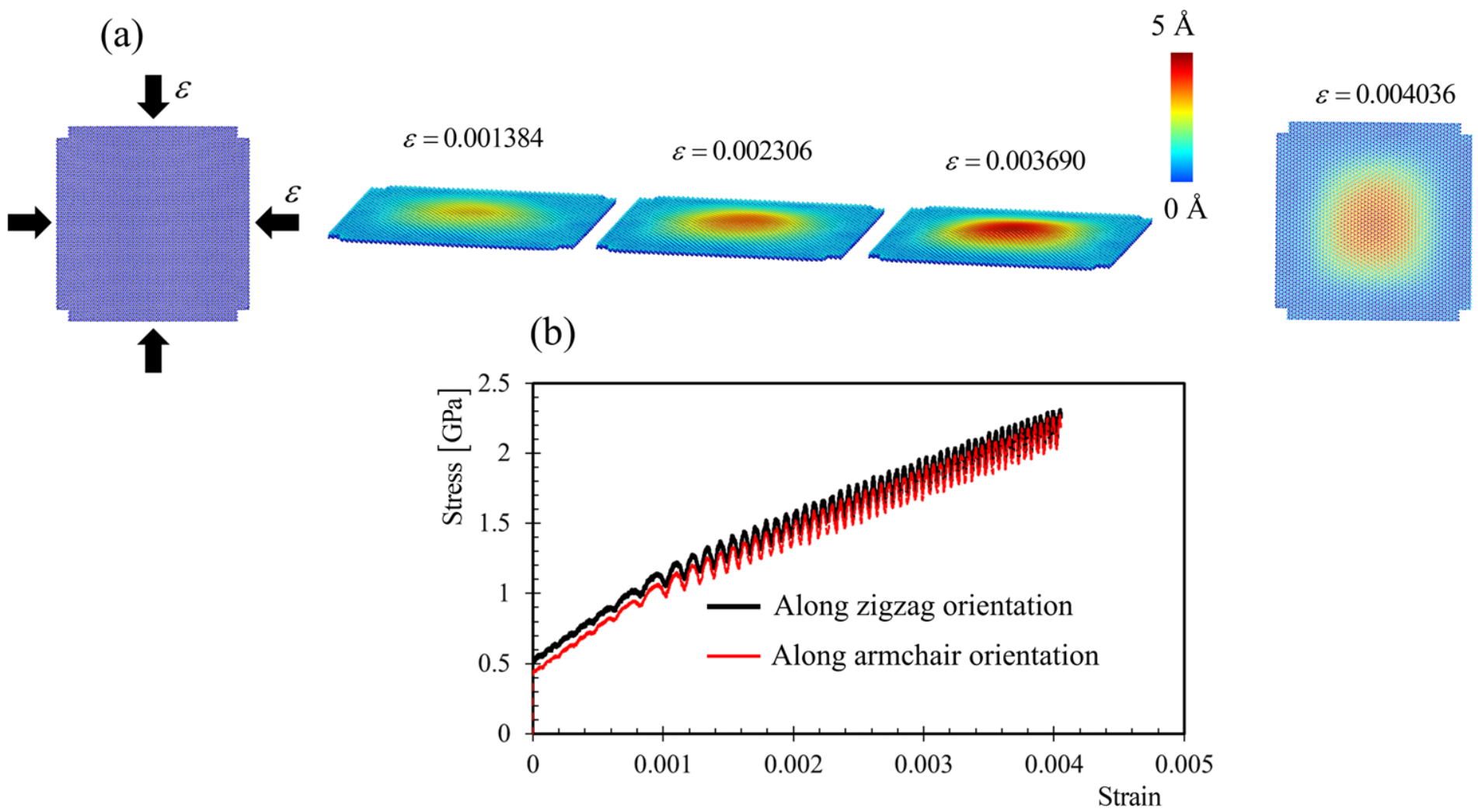

**Fig. 17** (a) Buckling mode shapes of a $Ti_2C$ MXene nanosheet with an in-plane dimension of $14\times14$ $nm^2$ under a biaxial compression with the strain rate of $1.45\times10^{6}$ $s^{-1}$ and timestep of 0.2 fs. Color maps represent the out-of-plane deflections. (b) Stress-strain curves along the armchair and the zigzag directions.

The response of the same $Ti_2C$ MXene nanosheet under pure shear deformation is presented in Fig. 18. Shear is applied by displacing the atoms in the manner illustrated in the figure, using a velocity of $5\times10^{-7}$ Å/fs. Under this loading, the iso-deformation contours form nearly concentric ellipses aligned with the shear direction. As the imposed shear increases, the nanosheet exhibits progressively larger out-of-plane deflection.

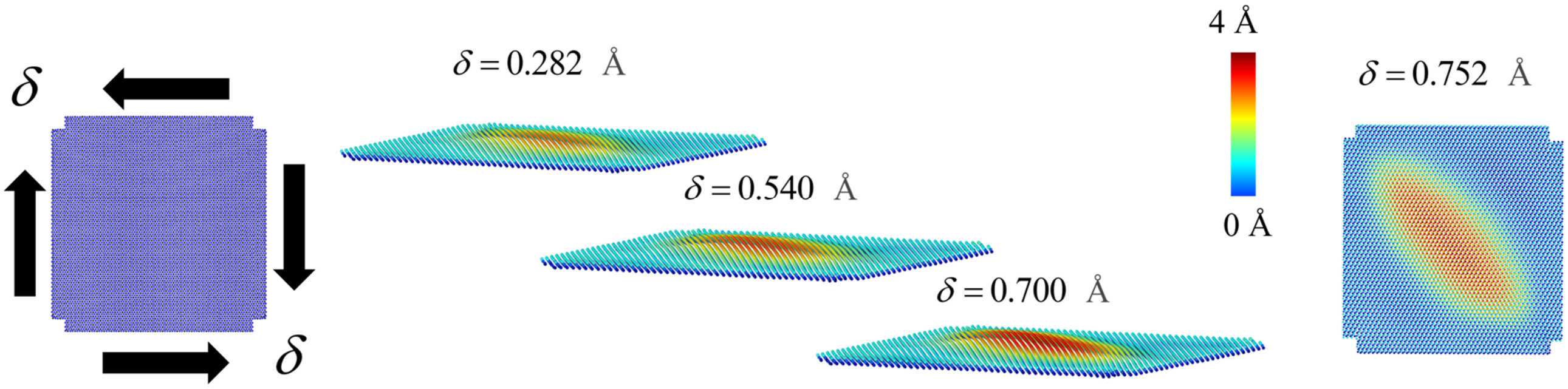

**Fig. 18** Buckling mode shapes of a $Ti_2C$ MXene nanosheet with an in-plane dimension of 14×14 $nm^2$ under a shear loading. The displacement δ is applied with a velocity of $5\times10^{-7}$ Å/fs, and the timestep is 0.2 fs. Color maps represent the out-of-plane deflections.

### 3.8. Combined Effect

So far, the individual effects of defects, lateral confinement, oxygen surface termination, and loading modes on the buckling behavior of $Ti_2C$ MXene nanosheets have been examined separately. In this section, we investigate the buckling response of the MXene under biaxial compression by considering the simultaneous presence of defects, oxygen surface termination, and lateral confinement.

The results, presented in Fig. 19, correspond to a defective $Ti_2CO_2$ MXene nanosheet with in-plane dimensions of 14 × 14 nm² subjected to biaxial compression at a strain rate of $5.8\times10^6$ $s^{-1}$ . The results include both the stress–strain response and the evolution of the buckling mode shape, where the color maps represent the out-of-plane deflections. During compression, the nanosheet is subjected to a confinement pressure modeled by two layers of harmonic springs positioned above and below the sheet, with an initial gap of 0.1 Å. The spring stiffness is defined as $k$ = 0.001 (kcal/mol)/Angstrom$^2$.

The nanosheet contains 1% isolated point defects for each atomic type, along with an additional 2% of Ti clustered defects, each cluster consisting of four vacancies. The defective atomic configuration was generated from the pristine MXene by selectively removing target atoms. Using a three-dimensional neighbor mapping with a bonding radius of 2.8 Å, any surface oxygen atom that lost its full titanium coordination due to a nearby Ti cluster vacancy was automatically removed. The resulting oxygen vacancies were then accounted for to ensure consistency with the prescribed 1% oxygen defect fraction.

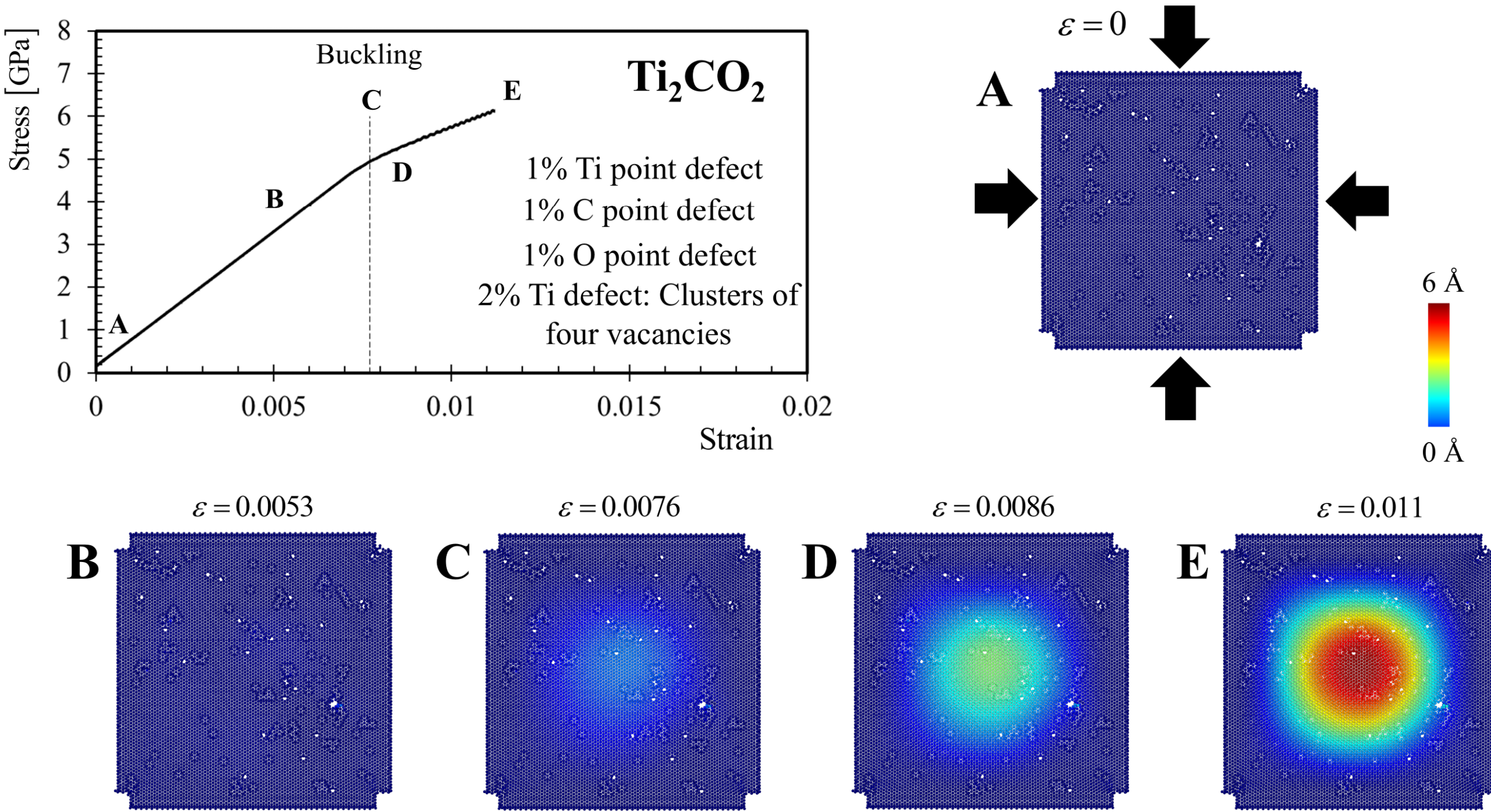

**Fig. 19** Stress-strain curve and buckling mode shape of a defective $Ti_2CO_2$ MXene nanosheet with an in-plane dimension of 14×14 nm$^2$ under a biaxial compression with the strain rate of 5.8×10$^6$ s$^{-1}$ and timestep of 0.2 fs. Color maps represent the out-of-plane deflections. The MXene nanosheet is under a confinement pressure introduced by positioning two layers of harmonic springs above and below the nanosheet with an initial gap of 0.1 Å. The spring stiffness is $k$ = 0.001 (kcal/mol)/Angstrom$^2$.

The stress initially increases linearly with strain up to point C (see figure), indicating a pre-buckling regime in which the MXene nanosheet retains its planar configuration. At a strain of approximately 0.0076 (point C), buckling is initiated, with the maximum out-of-plane deflection occurring near the center of the nanosheet, consistent with the behavior observed for $Ti_2C$ nanosheets under biaxial compression in Fig. 17. Beyond this point, the stress–strain response exhibits a reduced stiffness, accompanied by a progressive amplification of the buckling deformations.

### 3.9. Nonlocal Modeling

The effective thickness of a $Ti_2C$ MXene nanosheet is approximately 0.231 nm based on first-principles calculations reported in [56]. Applying Euler's buckling formula for a clamped–clamped beam of thickness $h$ = 0.231 nm and length $L$ = 14 nm yields a critical buckling strain of $\varepsilon_{buckling} = \left(\pi^2/3\right)\left(h/L\right)^2 = 0.0009$. Comparison with the buckling strain obtained from MD simulations in Fig. 4 (0.0015 for armchair-oriented loading) shows that the classical continuum mechanics prediction underestimates the buckling strain, highlighting its inadequacy for accurately describing the buckling behavior of $Ti_2C$ MXene nanosheets (note that the use of high strain rates in MD simulations may contribute to the observed discrepancy). This comparison further demonstrates that the $Ti_2C$ MXene nanosheet exhibits a stiffer response than predicted by the classical beam formulation, indicating that a size-dependent formulation is required to capture the buckling strain of the investigated nanosheets accurately [63–65]. Such a model has been previously developed, for instance, by the author of the present study in Ref. [66] based on the stress-driven nonlocal theory.

The nonlocal constitutive equation of a Bernoulli-Euler beam based on the stress-driven theory is given as [66–68]:

$$\chi - L_C^2 \frac{d^2 \chi}{dx^2} = \frac{M}{EI} \tag{1}$$

where $\chi(x)$ is the curvature with $0 \le x \le L$ and $L$ the length of the beam, $M(x)$ the bending moment, $E$ the Young modulus, and $I$ the second moment of area. The characteristic length parameter, $L_C$, is related to microstructure properties and can be calibrated using MD [69] or experimental results [70]. The nonlocal constitutive equation (1) reduces to the classical Bernoulli-Euler beam formulation for $L_C = 0$.

The stress-driven constitutive equation (1) is of second order. The mathematical well-posedness is guaranteed by the following constitutive boundary conditions at the beam ends [66–68]:

$$\begin{aligned} L_C \frac{d\chi(0)}{dx} - \chi(0) = 0 \\ L_C \frac{d\chi(L)}{dx} + \chi(L) = 0 \end{aligned} \tag{2}$$

The buckling load and the corresponding critical strain are determined by solving the eigenvalue problem derived from the variationally consistent governing equation, subject to clamped–clamped boundary conditions and the higher-order constitutive boundary conditions defined in Eq. (2). For brevity, the explicit forms of these equations are not reported here and can be found in Refs. [66,68].

The MD predictions of the buckling strains of $Ti_2C$ MXene nanosheets with in-plane dimensions of $L \times L$, subjected to uniaxial compression along the armchair direction at a strain rate of $1.45\times10^{6}$ $s^{-1}$ and timestep of 0.2 fs are shown in Fig. 20 by circle markers. The results correspond to the nanosheets with $L$ = 4, 6, 8, 10, 12, and 14 nm. The results are reported for a range of in-plane dimensions, thereby enabling a systematic assessment of size-dependent behavior. A clear dependence of the buckling strain on the nanosheet size is observed, different than that predicted by the classical beam formulation, $\varepsilon_{buckling} \propto L^{-2}$. In contrast, the stress-driven nonlocal model, incorporating an appropriate characteristic length parameter, $L_C$, can reproduce the MD predictions with good accuracy across the considered range of dimensions. By increasing the value of the characteristic length parameter, $L_C$, the predictions of the stress-driven formulation deviate more from the classical beam theory. For $L_C = 5$ Å, the predictions of the stress-driven model capture the MD simulation results with good accuracy. The predicted characteristic length parameter is of the same order of magnitude as the first-principles predictions of the $Ti_2C$ MXene lattice parameter and Ti–C bond length, 3.035 Å and 2.097 Å [58].

This agreement highlights the capability of the nonlocal formulation to capture the underlying mechanics governing the buckling response of MXene nanosheets.

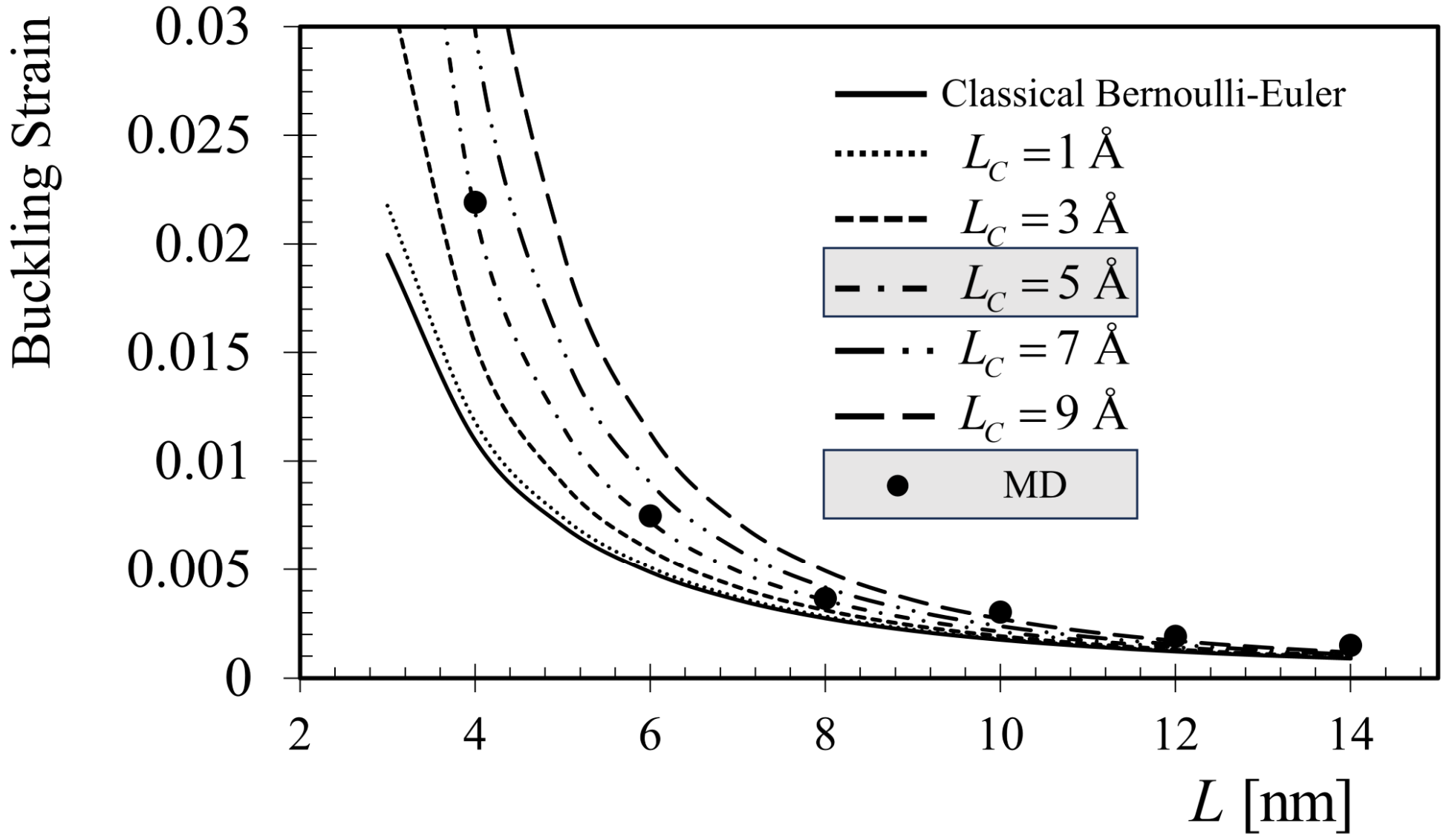

**Fig. 20** Buckling strains of $Ti_2C$ MXene nanosheets with an in-plane dimensions of $L \times L$ under compression along the armchair direction at a strain rate of $1.45 \times 10^6$ s$^{-1}$ and timestep of 0.2 fs. Results are presented for different in-plane dimensions. The stress-driven nonlocal model with the characteristic length parameter $L_C = 5$ Å captures the MD results with good accuracy.

## 4. Conclusions

A comprehensive atomistic investigation into the compressive and post-buckling behavior of $Ti_2C$ and $Ti_2CO_2$ MXene nanosheets has been presented in this study using reactive molecular dynamics simulations. The results demonstrate that classical continuum mechanics significantly underestimates the buckling strains of 2D MXene nanosheets. A previously developed stress-driven nonlocal theory is then employed to show that this class of nonlocal formulations can accurately capture the size-dependent buckling response of the investigated nanosheets.

$Ti_2C$ nanosheets exhibit greater resistance to buckling along the armchair direction than along the zigzag direction. Higher strain rates shift the onset of buckling to higher stress and strain levels and induce a more unstable post-buckling response characterized by larger stress oscillations. While isolated point defects (up to 3%) reduce buckling loads, vacancy clusters fundamentally change the global buckling deformation. For clusters consisting of four or five titanium vacancies, the compressive response of MXene deviates from traditional thin-plate buckling behavior; rather, a wavy deformation pattern driven by the severe localized compliance around the defects is formed. Surface termination has a pronounced effect on compressive stability: oxygen termination substantially increases the buckling stress and strain, and reduces directional anisotropy in the elastic response. At large compressive strains, $Ti_2CO_2$ nanosheets undergo fracture, whereas $Ti_2C$ nanosheets retain structural integrity even beyond strains of 0.35. The simulations further show that lateral confinement pressure strongly enhances buckling resistance and mitigates post-buckling oscillations. Under biaxial compression, the nanosheets buckle into a dome-like shape, while shear loading yields elliptical deflection modes. We have also demonstrated that the developed MD simulation framework can model the complex, synergistic effects of defects, surface chemistry, confinement, and varying loading types on the buckling response of MXene nanosheets.

The findings in this paper highlight the complex interplay among surface chemistry, defect structure, and loading conditions in governing the mechanical stability of MXene nanosheets. These insights provide both a fundamental understanding and practical guidance for the design and optimization of MXene-based composites, flexible electronics, and nanoscale structural components in which compressive stability is essential. Additionally, the present results can stimulate future explorations of MXene morphological transformations, including the emergence of nanotube, nanoscroll, and folded architectures.

## Declaration of interests

The author declares that he has no known competing financial interests or personal relationships that could have appeared to influence the work reported in this paper. Hossein Darban is an Editorial Board Member of this journal and was not involved in the editorial review or the decision to publish this article.

## Funding

The author gratefully acknowledges the financial support provided by the National Science Centre (NCN) in Poland through the grant agreement No: UMO-2022/47/D/ST8/01348.

## Acknowledgements

The author gratefully acknowledges the Polish high-performance computing infrastructure PLGrid (HPC Centers: WCSS, ACK Cyfronet AGH) for providing computer facilities and support within the computational grant no. PLG/2026/019291.

## References

[1] M. Naguib, M. Kurtoglu, V. Presser, J. Lu, J. Niu, M. Heon, L. Hultman, Y. Gogotsi, M.W. Barsoum, Two-Dimensional Nanocrystals Produced by Exfoliation of Ti3AlC2, Advanced Materials 23 (2011) 4248–4253. https://doi.org/10.1002/adma.201102306.

[2] G. Murali, J.K. Reddy Modigunta, Y.H. Park, J.-H. Lee, J. Rawal, S.-Y. Lee, I. In, S.-J. Park, A Review on MXene Synthesis, Stability, and Photocatalytic Applications, ACS Nano 16 (2022) 13370–13429. https://doi.org/10.1021/acsnano.2c04750.

[3] G. Murali, J. Rawal, J.K.R. Modigunta, Y.H. Park, J.-H. Lee, S.-Y. Lee, S.-J. Park, I. In, A review on MXenes: new-generation 2D materials for supercapacitors, Sustainable Energy Fuels 5 (2021) 5672–5693. https://doi.org/10.1039/D1SE00918D.

[4] Y. Zhang, Z. Zhou, J. Lan, P. Zhang, Prediction of Ti3C2O2 MXene as an effective capturer of formaldehyde, Applied Surface Science 469 (2019) 770–774. https://doi.org/10.1016/j.apsusc.2018.11.018.

[5] Z. Shi, S. Zhong, Z. Zhang, Y. Sui, J. Zhou, L. Wu, Atomic insights into monolayer MXene degradation and impacts for lithium-ion storage, Applied Surface Science 689 (2025) 162506. https://doi.org/10.1016/j.apsusc.2025.162506.

[6] S.M. Hatam-Lee, A. Esfandiar, A. Rajabpour, Mechanical behaviors of titanium nitride and carbide MXenes: A molecular dynamics study, Applied Surface Science 566 (2021) 150633. https://doi.org/10.1016/j.apsusc.2021.150633.

[7] A. VahidMohammadi, J. Rosen, Y. Gogotsi, The world of two-dimensional carbides and nitrides (MXenes), Science 372 (2021) eabf1581. https://doi.org/10.1126/science.abf1581.

[8] X. Xu, T. Guo, M. Lanza, H.N. Alshareef, Status and prospects of MXene-based nanoelectronic devices, Matter 6 (2023) 800–837. https://doi.org/10.1016/j.matt.2023.01.019.

[9] F. Xie, Q. Liu, L. Zhuo, H. Wei, Y. Shang, T. Liu, Z. Lu, Hierarchically engineered Sandwich-Structured h-MXene/ANF hybrid films with tunable electromagnetic interference shielding and exceptional environmental resilience, Composite Structures 367 (2025) 119248. https://doi.org/10.1016/j.compstruct.2025.119248.

[10] J. Zheng, Y. Deng, D. Luo, F. Wu, X. Dai, Preparation and performance enhancement of n-eicosane/polyvinyl alcohol/MXene flexible phase change composites with sandwich structure, Composite Structures 331 (2024) 117930. https://doi.org/10.1016/j.compstruct.2024.117930.

[11] H. Cao, Beyond graphene and boron nitride: why MXene can be used in composite for corrosion protection on metals?, Composites Part B: Engineering 271 (2024) 111168. https://doi.org/10.1016/j.compositesb.2023.111168.

[12] Z. Lu, X. Shen, Z. Li, Q. Jia, Flexible sandwich structure MXene-Aramid Nanofiber-MXene film for adjusting infrared camouflage and multifunctional application, Composite Structures 343 (2024) 118299. https://doi.org/10.1016/j.compstruct.2024.118299.

[13] T.T. Huynh, H.Q. Pham, Defect and non-metallic doping co-engineering in two-dimensional transition-metal carbide-based electrocatalysts for renewable energy conversion, Coordination Chemistry Reviews 549 (2026) 217254. https://doi.org/10.1016/j.ccr.2025.217254.

[14] Y. Liu, D. Lv, Y. Yang, L. Zhu, R. Wang, S. Chen, D. Jiang, H. Wang, Multi-dimensional applications of MXene in human health: From disease prevention, diagnosis, to rehabilitation, Journal of Materials Science & Technology 246 (2026) 187–219. https://doi.org/10.1016/j.jmst.2025.02.086.

[15] Y. Yao, X. Li, K.M. Sisican, R.M.C. Ramos, M. Judicpa, S. Qin, J. Zhang, J. Yao, J.M. Razal, K.A.S. Usman, Progress towards efficient MXene sensors, Commun Mater 6 (2025) 210. https://doi.org/10.1038/s43246-025-00907-y.

[16] G. Singh, A. Kaur, J.W. Lim, MXene-SnO2 modified carbon felt electrode for enhanced performance in vanadium redox flow batteries, Composite Structures 379 (2026) 120001. https://doi.org/10.1016/j.compstruct.2025.120001.

[17] Y. Zhang, X. Wu, Q. Guo, D. Zhang, C. Li, D. Li, Y. Liu, J. Zhang, P. Zhang, Y. Yan, L. An, J. Guo, L. Chen, Advances in sensors technologies for composites structural health monitoring, Composite Structures 370 (2025) 119448. https://doi.org/10.1016/j.compstruct.2025.119448.

[18] I. Amin, H. van den Brekel, K. Nemani, E. Batyrev, A. de Vooys, H. van der Weijde, B. Anasori, N.R. Shiju, Ti3C2Tx MXene Polymer Composites for Anticorrosion: An Overview and Perspective, ACS Appl. Mater. Interfaces 14 (2022) 43749–43758. https://doi.org/10.1021/acsami.2c11953.

[19] L. Kabir, K. Wijaya, S. Sagadevan, J. Unruangsri, K. Ullah, W.-C. Oh, 2D MXene-polymer based composites for biomedical applications, Journal of Alloys and Compounds 1031 (2025) 181068. https://doi.org/10.1016/j.jallcom.2025.181068.

[20] A. Lipatov, H. Lu, M. Alhabeb, B. Anasori, A. Gruverman, Y. Gogotsi, A. Sinitskii, Elastic properties of 2D Ti3C2Tx MXene monolayers and bilayers, Science Advances 4 (2018) eaat0491. https://doi.org/10.1126/sciadv.aat0491.

[21] J.W. Suk, R.D. Piner, J. An, R.S. Ruoff, Mechanical Properties of Monolayer Graphene Oxide, ACS Nano 4 (2010) 6557–6564. https://doi.org/10.1021/nn101781v.

[22] C. Gómez-Navarro, M. Burghard, K. Kern, Elastic Properties of Chemically Derived Single Graphene Sheets, Nano Lett. 8 (2008) 2045–2049. https://doi.org/10.1021/nl801384y.

[23] C. Rong, T. Su, Z. Li, T. Chu, M. Zhu, Y. Yan, B. Zhang, F.-Z. Xuan, Elastic properties and tensile strength of 2D Ti3C2Tx MXene monolayers, Nat Commun 15 (2024) 1566. https://doi.org/10.1038/s41467-024-45657-6.

[24] Z. Guo, J. Zhou, C. Si, Z. Sun, Flexible two-dimensional Tin+1Cn (n = 1, 2 and 3) and their functionalized MXenes predicted by density functional theories, Phys. Chem. Chem. Phys. 17 (2015) 15348–15354. https://doi.org/10.1039/C5CP00775E.

[25] J.D. Gouveia, T.L.P. Galvão, K. Iben Nassar, J.R.B. Gomes, First-principles and machine-learning approaches for interpreting and predicting the properties of MXenes, Npj 2D Mater Appl 9 (2025) 8. https://doi.org/10.1038/s41699-025-00529-5.

[26] K. Chenoweth, A.C.T. van Duin, W.A. Goddard, ReaxFF Reactive Force Field for Molecular Dynamics Simulations of Hydrocarbon Oxidation, J. Phys. Chem. A 112 (2008) 1040–1053. https://doi.org/10.1021/jp709896w.

[27] Q. Yang, S.J. Eder, A. Martini, P.G. Grützmacher, Effect of surface termination on the balance between friction and failure of Ti3C2Tx MXenes, Npj Mater Degrad 7 (2023) 6. https://doi.org/10.1038/s41529-023-00326-9.

[28] I. Greenfeld, S. Jiang, L. Yang, H.D. Wagner, Nonlinear elasticity degrades monolayer fracture toughness, Acta Materialia 286 (2025) 120727. https://doi.org/10.1016/j.actamat.2025.120727.

[29] G. Plummer, B. Anasori, Y. Gogotsi, G.J. Tucker, Nanoindentation of monolayer Ti*n+1*C*n*T*x* MXenes via atomistic simulations: The role of composition and defects on strength, Computational Materials Science 157 (2019) 168–174. https://doi.org/10.1016/j.commatsci.2018.10.033.

[30] Y.I. Jhon, Y.T. Byun, J.H. Lee, Y.M. Jhon, Robust mechanical tunability of 2D transition metal carbides via surface termination engineering: Molecular dynamics simulation, Applied Surface Science 532 (2020) 147380. https://doi.org/10.1016/j.apsusc.2020.147380.

[31] C. Wei, C. Wu, Nonlinear fracture of two-dimensional transition metal carbides (MXenes), Engineering Fracture Mechanics 230 (2020) 106978. https://doi.org/10.1016/j.engfracmech.2020.106978.

[32] H. Yu, K. Xu, Z. Zhang, X. Cao, J. Weng, J. Wu, Oxygen functionalization-induced crossover in the tensile properties of the thinnest 2D Ti2C MXene, J. Mater. Chem. C 9 (2021) 2416–2425. https://doi.org/10.1039/D0TC04637J.

[33] N. Nayir, Q. Mao, T. Wang, M. Kowalik, Y. Zhang, M. Wang, S. Dwivedi, G.-U. Jeong, Y.K. Shin, A. van Duin, Modeling and simulations for 2D materials: a ReaxFF perspective, 2D Mater. 10 (2023) 032002. https://doi.org/10.1088/2053-1583/acd7fd.

[34] Y. Chen, S. Tang, X. Yan, Manipulating the crack path through the surface functional groups of MXenes, Nanoscale 14 (2022) 14169–14177. https://doi.org/10.1039/D2NR02235D.

[35] K. Ganeshan, Y.K. Shin, N.C. Osti, Y. Sun, K. Prenger, M. Naguib, M. Tyagi, E. Mamontov, D. Jiang, A.C.T. van Duin, Structure and Dynamics of Aqueous Electrolytes Confined in 2D-TiO2/Ti3C2T2

MXene Heterostructures, ACS Appl. Mater. Interfaces 12 (2020) 58378–58389. https://doi.org/10.1021/acsami.0c17536.

[36] N.C. Osti, M. Naguib, A. Ostadhossein, Y. Xie, P.R.C. Kent, B. Dyatkin, G. Rother, W.T. Heller, A.C.T. van Duin, Y. Gogotsi, E. Mamontov, Effect of Metal Ion Intercalation on the Structure of MXene and Water Dynamics on its Internal Surfaces, ACS Appl. Mater. Interfaces 8 (2016) 8859–8863. https://doi.org/10.1021/acsami.6b01490.

[37] L.F.V. Thomazini, A.F. Fonseca, Structure and elastic properties of titanium MXenes: Evaluation of COMB3, REAXFF and MEAM force fields, Computational Materials Science 259 (2025) 114134. https://doi.org/10.1016/j.commatsci.2025.114134.

[38] Md.M. Billah, M.S. Rabbi, K.A. Rahman, P. Acar, Temperature and strain rate dependent tensile properties of titanium carbide/nitride MXenes, Materials Chemistry and Physics 312 (2024) 128581. https://doi.org/10.1016/j.matchemphys.2023.128581.

[39] V.N. Borysiuk, V.N. Mochalin, Y. Gogotsi, Molecular dynamic study of the mechanical properties of two-dimensional titanium carbides Tin+1Cn (MXenes), Nanotechnology 26 (2015) 265705. https://doi.org/10.1088/0957-4484/26/26/265705.

[40] A.H. Khadem, H.R. Anik, M.G.M. Limon, M. Rabbani, A.S. Rafi, I.K. Mim, M.H. Rahman, F.B. Alam, M.K. Islam, S.S. Munira, M.T. Islam, M. Uddin, H. Mony, Review of MXene-Based Electrospun Nanocomposites for Flexible and Wearable Sensors, ACS Appl. Nano Mater. 8 (2025) 16231–16259. https://doi.org/10.1021/acsanm.5c02646.

[41] M. Xin, J. Li, Z. Ma, L. Pan, Y. Shi, MXenes and Their Applications in Wearable Sensors, Front. Chem. 8 (2020). https://doi.org/10.3389/fchem.2020.00297.

[42] L. Zhao, K. Wang, W. Wei, L. Wang, W. Han, High-performance flexible sensing devices based on polyaniline/MXene nanocomposites, InfoMat 1 (2019) 407–416. https://doi.org/10.1002/inf2.12032.

[43] A.K. Katiyar, A.T. Hoang, D. Xu, J. Hong, B.J. Kim, S. Ji, J.-H. Ahn, 2D Materials in Flexible Electronics: Recent Advances and Future Prospectives, Chem. Rev. 124 (2024) 318–419. https://doi.org/10.1021/acs.chemrev.3c00302.

[44] O. Lourie, D.M. Cox, H.D. Wagner, Buckling and Collapse of Embedded Carbon Nanotubes, Phys. Rev. Lett. 81 (1998) 1638–1641. https://doi.org/10.1103/PhysRevLett.81.1638.

[45] Y. Li, C. Wei, S. Huang, C. Wu, V.N. Mochalin, *In-situ* SEM compression of accordion-like multilayer MXenes, Extreme Mechanics Letters 41 (2020) 101054. https://doi.org/10.1016/j.eml.2020.101054.

[46] A. Genoese, A. Genoese, N.L. Rizzi, G. Salerno, Buckling Analysis of Single-Layer Graphene Sheets Using Molecular Mechanics, Front. Mater. 6 (2019). https://doi.org/10.3389/fmats.2019.00026.

[47] A. Montazeri, S. Ebrahimi, H. Rafii-Tabar, A molecular dynamics investigation of buckling behaviour of hydrogenated graphene, Molecular Simulation 41 (2015) 1212–1218. https://doi.org/10.1080/08927022.2014.968849.

[48] A.P. Sgouros, G. Kalosakas, K. Papagelis, C. Galiotis, Compressive response and buckling of graphene nanoribbons, Sci Rep 8 (2018) 9593. https://doi.org/10.1038/s41598-018-27808-0.

[49] Y. Xiang, H.-S. Shen, Shear buckling of rippled graphene by molecular dynamics simulation, Materials Today Communications 3 (2015) 149–155. https://doi.org/10.1016/j.mtcomm.2015.01.001.

[50] S. Plimpton, Fast Parallel Algorithms for Short-Range Molecular Dynamics, Journal of Computational Physics 117 (1995) 1–19. https://doi.org/10.1006/jcph.1995.1039.

[51] J.R. Gissinger, I. Nikiforov, Y. Afshar, B. Waters, M. Choi, D.S. Karls, A. Stukowski, W. Im, H. Heinz, A. Kohlmeyer, E.B. Tadmor, Type Label Framework for Bonded Force Fields in LAMMPS, J. Phys. Chem. B 128 (2024) 3282–3297. https://doi.org/10.1021/acs.jpcb.3c08419.

[52] H.M. Aktulga, J.C. Fogarty, S.A. Pandit, A.Y. Grama, Parallel reactive molecular dynamics: Numerical methods and algorithmic techniques, Parallel Computing 38 (2012) 245–259. https://doi.org/10.1016/j.parco.2011.08.005.

[53] A. Stukowski, Visualization and analysis of atomistic simulation data with OVITO–the Open Visualization Tool, Modelling Simul. Mater. Sci. Eng. 18 (2009) 015012. https://doi.org/10.1088/0965-0393/18/1/015012.

[54] M. Khazaei, M. Arai, T. Sasaki, C.-Y. Chung, N.S. Venkataramanan, M. Estili, Y. Sakka, Y. Kawazoe, Novel Electronic and Magnetic Properties of Two-Dimensional Transition Metal Carbides and Nitrides, Advanced Functional Materials 23 (2013) 2185–2192. https://doi.org/10.1002/adfm.201202502.

[55] W.C. Swope, H.C. Andersen, P.H. Berens, K.R. Wilson, A computer simulation method for the calculation of equilibrium constants for the formation of physical clusters of molecules: Application to small water clusters, The Journal of Chemical Physics 76 (1982) 637–649. https://doi.org/10.1063/1.442716.

[56] T. Hu, J. Yang, W. Li, X. Wang, C.M. Li, Quantifying the rigidity of 2D carbides (MXenes), Phys. Chem. Chem. Phys. 22 (2020) 2115–2121. https://doi.org/10.1039/C9CP05412J.

[57] H. Darban, Auxetic Response in Two-Dimensional MXenes with Atomically Defined Perforations, (2026). https://doi.org/10.48550/arXiv.2603.19436.

[58] Y. Bai, K. Zhou, N. Srikanth, J.H.L. Pang, X. He, R. Wang, Dependence of elastic and optical properties on surface terminated groups in two-dimensional MXene monolayers: a first-principles study, RSC Adv. 6 (2016) 35731–35739. https://doi.org/10.1039/C6RA03090D.

[59] L.-Å. Näslund, M. Magnuson, The origin of Ti 1s XANES main edge shifts and EXAFS oscillations in the energy storage materials Ti2CTx and Ti3C2Tx MXenes, 2D Mater. 10 (2023) 035024. https://doi.org/10.1088/2053-1583/acd7fe.

[60] X. Sang, Y. Xie, M.-W. Lin, M. Alhabeb, K.L. Van Aken, Y. Gogotsi, P.R.C. Kent, K. Xiao, R.R. Unocic, Atomic Defects in Monolayer Titanium Carbide (Ti3C2Tx) MXene, ACS Nano 10 (2016) 9193–9200. https://doi.org/10.1021/acsnano.6b05240.

[61] Y. Lanir, Y.C.B. Fung, Fiber Composite Columns Under Compression, Journal of Composite Materials 6 (1972) 387–401. https://doi.org/10.1177/002199837200600315.

[62] X.-T. Dang, V.-L. Nguyen, M.-T. Tran, B.-P. Nguyen-Thi, T.-T. Le, Buckling and bending analysis of FGP nanoplates resting on Pasternak foundation considering non-local and surface effects

simultaneously using pb2-Ritz method, Composite Structures 359 (2025) 118971. https://doi.org/10.1016/j.compstruct.2025.118971.

[63] B. Yang, N. Fantuzzi, M. Bacciocchi, F. Fabbrocino, M. Mousavi, Dynamic characteristics of multilayer graphene considering nonlinear higher-order strain gradient, Composite Structures 376 (2026) 119800. https://doi.org/10.1016/j.compstruct.2025.119800.

[64] B. Yang, N. Fantuzzi, M. Bacciocchi, F. Fabbrocino, M. Mousavi, Nonlinear wave propagation in graphene incorporating second strain gradient theory, Thin-Walled Structures 198 (2024) 111713. https://doi.org/10.1016/j.tws.2024.111713.

[65] B. Yang, M. Mousavi, Nonlinear dynamic behavior of carbon nanotubes incorporating size effects, International Journal of Mechanical Sciences 267 (2024) 109014. https://doi.org/10.1016/j.ijmecsci.2024.109014.

[66] H. Darban, F. Fabbrocino, L. Feo, R. Luciano, Size-dependent buckling analysis of nanobeams resting on two-parameter elastic foundation through stress-driven nonlocal elasticity model, Mechanics of Advanced Materials and Structures 28 (2021) 2408–2416. https://doi.org/10.1080/15376494.2020.1739357.

[67] G. Romano, R. Barretta, Nonlocal elasticity in nanobeams: the stress-driven integral model, International Journal of Engineering Science 115 (2017) 14–27. https://doi.org/10.1016/j.ijengsci.2017.03.002.

[68] R. Barretta, F. Fabbrocino, R. Luciano, F.M. de Sciarra, G. Ruta, Buckling loads of nano-beams in stress-driven nonlocal elasticity, Mechanics of Advanced Materials and Structures 27 (2020) 869–875. https://doi.org/10.1080/15376494.2018.1501523.

[69] H. Darban, MD benchmarks: Size-dependent tension, bending, buckling, and vibration of nanobeams, International Journal of Mechanical Sciences 296 (2025) 110316. https://doi.org/10.1016/j.ijmecsci.2025.110316.

[70] H. Darban, R. Luciano, M. Basista, Calibration of the length scale parameter for the stress-driven nonlocal elasticity model from quasi-static and dynamic experiments, Mechanics of Advanced Materials and Structures 30 (2023) 3518–3524. https://doi.org/10.1080/15376494.2022.2077488.